\documentclass[aps,prd,twocolumn,superscriptaddress]{revtex4-2} \usepackage{lipsum}
\usepackage{graphicx}  
\usepackage{dcolumn}   
\usepackage{bm}        
\usepackage{amssymb}   
\usepackage{epsfig}
\usepackage{epstopdf}  
\usepackage{subfigure}
\usepackage{mdframed}
\usepackage{amsmath}

\usepackage[colorlinks=true]{hyperref}  
\hypersetup{
    bookmarks=true,         
    unicode=false,          
    pdftoolbar=true,        
    pdfmenubar=true,        
    pdffitwindow=false,     
    pdfstartview={FitH},    
    pdftitle={My title},    
    pdfauthor={Author},     
    pdfsubject={Subject},   
    pdfcreator={Creator},   
    pdfproducer={Producer}, 
    pdfkeywords={keyword1} {key2} {key3}, 
    pdfnewwindow=true,      
    colorlinks=true,       
    linkcolor=magenta, 
    citecolor=blue,        
    filecolor=magenta,      
    urlcolor=cyan           
} 
\newcommand{\ssc}{\scriptscriptstyle}
\newcommand\eea{\end{eqnarray}}
\newcommand\bea{\begin{eqnarray}}
\newcommand{\eg}{{\it e.g.,}\ }
\newcommand{\ie}{{\it i.e.,}\ }

\newcommand{\diff}{\mathrm{d}}

\newcommand{\be}{\begin{equation}}
\newcommand{\ee}{\end{equation}}
\newcommand{\ba}{\begin{align}}
\newcommand{\ea}{\end{align}}
\newcommand{\bg}{\begin{gather}}

\usepackage{graphicx}
\usepackage{dcolumn}
\usepackage{bm}

\begin{document}


\title{Generalized Symmetries For Generalized Gravitons}

\author{Valentin Benedetti}
 \email{valentin.benedetti@ib.edu.ar}
\affiliation{%
Instituto Balseiro, Centro At\'omico Bariloche\\
 8400-S.C. de Bariloche, R\'io Negro, Argentina
}%

\author{Pablo Bueno}
 \email{pablobueno@ub.edu}
\affiliation{%
Departament de F\'isica Qu\`antica i Astrof\'isica, Institut de Ci\`encies del Cosmos\\
 Universitat de Barcelona, Mart\'i i Franqu\`es 1, E-08028 Barcelona, Spain 
} %


\author{Javier M. Magan}
 \email{javier.magan@cab.cnea.gov.ar}
\affiliation{%
Instituto Balseiro, Centro At\'omico Bariloche\\
 8400-S.C. de Bariloche, R\'io Negro, Argentina
}%


\date{\today}

\begin{abstract}
We construct generalized symmetries for linearized Einstein gravity in arbitrary dimensions.
First-principle considerations in QFT force generalized symmetries to appear in dual pairs. Verifying this prediction helps us find the full set of non-trivial conserved charges ---associated, in equal parts, with 2-form and $(D-2)$-form currents. Their total number is $D(D+1)$. 
We compute the quantum commutators of pairs of dual charges, showing that they are non-vanishing for regions  whose boundaries are non-trivially linked with each other and zero otherwise, as expected on general grounds.  
We also consider  general linearized higher-curvature gravities. These propagate, in addition to the usual graviton, a spin-0 mode as well as a massive ghost-like spin-2 one. When the latter is absent, the theory is unitary and the dual-pairs principle is respected. In particular, we find that the number and types of charges remain the same as for Einstein gravity, and that they correspond to continuous generalizations of the Einsteinian ones. 


\end{abstract}

\maketitle


The notion of generalized symmetry can be traced back to 't Hooft's seminal  paper \cite{tHooft:1977nqb}, but it has been recently put on a broader and firmer ground by Gaiotto et al. \cite{Gaiotto:2014kfa}. The idea is that symmetry operators can live in codimension-one/two/$\dots$ hypersurfaces. Dually, the  charged operators can be local/line/$\dots$ operators. In this context the Landau paradigm gets extended to include a far larger zoo of physical theories, most importantly gauge theories. Recent reviews include \cite{Cordova:2022ruw,McGreevy:2022oyu,Casini:2021zgr}. 

It is then important to inquire about the interplay between generalized symmetries and gravity. A conservative starting point is to consider gravitons (spin-two fields) on Minkoswki spacetime \cite{Fierz:1939ix}. The simplest 
case corresponds to the Einstein ---or Fierz-Pauli--- graviton. The generalized symmetries of this theory were studied recently in \cite{casini2021generalized,Benedetti:2022zbb,hofman}, mostly in $D=4$ spacetime dimensions. An insightful outcome is that generalized symmetries are charged under spacetime symmetries, shedding important new light on Noether's and Weinberg-Witten's theorems \cite{Benedetti:2022zbb}, and providing intriguing relations between the gravitational interaction and the physics of fractons \cite{Pretko:2017,casini2021generalized,hofman}. The next-to-simplest scenario is to go beyond $D=4$ and consider general theories of linearized gravity, dubbed here theories of ``generalized gravitons''. Characterizing the generalized symmetries of these theories is the main goal of this letter. 

There are several motivations for this analysis. First, the study of conserved charges in gravity theories has been a key area in the field, one of its highlights being the Wald formalism \cite{Lee:1990nz,Wald:1993nt,Iyer_1994}. If new conserved charges exist, it is important to find them. 
Second, as we review below, generalized symmetries in QFT always come in dual pairs \cite{Casini:2020rgj,Magan:2020ake,Casini:2021zgr}. 
This principle has important implications. 
In the holographic context, it provides an argument against the existence of higher-form symmetries in quantum gravity \cite{Casini:2021zgr}. 
It also lies behind the proof of the universal charged density of states in QFT \cite{Casini:2019kex,Magan:2021myk,Harlow:2021trr}. In certain scenarios, such as the ones considered in this letter, it might predict the existence of new, otherwise unexpected, charges. Conversely, it might suggest that certain naive charges do not generate new symmetries. 
Third, within quantum gravity the principle of the completeness of the spectrum in gauge theories \cite{Polchinski:2003bq,Banks:2010zn} has recently been connected to the absence of generalized symmetries \cite{Rudelius:2020orz,Casini:2020rgj,Heidenreich:2020tzg,Casini:2021zgr,Witten:2023qsv}. Studying the generalized symmetries of generalized gravitons we are advancing towards understanding what kind of matter is required to break these symmetries. In fact, for the Einstein graviton, Ref.\,\cite{casini2021generalized} showed that conventional ways of breaking the generalized symmetries lead to the breaking of Poincar\'e invariance. 
One last motivation comes from condensed matter physics and the connection between gravity and fractons \cite{Pretko:2020}. 
The zoo of theories considered here enlarges the space of potential fractonic systems as well as the one of tensor gauge theories \cite{Rasmussen2016}.

 
{\bf Generalized Symmetries, Algebras and Conserved Currents:}
In $D$-dimensional flat space, a $p$-form symmetry current $J$ is a $p$-form which satisfies the conservation property $\diff \star J=0$,
where $\star\, J$ is the $(D-p)$-form defined by the Hodge dual of $J$ and $\diff $ is the exterior derivative. Conserved currents like $J$ define conserved higher-form charges $Q$ by integrating $\star\, J$ over closed $(D-p)$ surfaces $\Sigma_{(D-p)}$ embedded in $\mathbb{R}^D$, namely \cite{Gaiotto:2014kfa}
\be 
\Phi= \int_{\Sigma_{(D-p)}} \star\, J\, .
\label{hfch1}
\ee
Given $\Phi$, the operator that implements the generalized symmetry reads $U_g=e^{{\rm i}\, g \, \Phi}$.  To have a true symmetry,  we need $\Phi\neq 0$ and this implies $\star\, J\,\neq\, \diff\,G$, where $G$ is a physical field of the theory. The paradigmatic example is the free Maxwell field, which has two conserved currents $F$ and $\star\, F$, where $F$ is the field strength. Although $F=\diff\,A$ and $\star \,F=\diff\,\tilde{A}$, neither $A$ nor $\tilde{A}$ are physical fields. 

An algebraic approach to the notion of generalized symmetry appeared in \cite{Casini:2020rgj,Casini:2021zgr}. A QFT naturally assigns von Neumann algebras of operators ${\cal A}(R)$ to spacetime regions $R$. 
A minimal assignation corresponds to the ``additive algebra'', ${\cal A}_{\textrm{add}}(R) $, {\it i.e.,} the algebra generated by local operators in $R$. Causality then forces
\be
{\cal A}_{\textrm{add}}(R)   \subseteq  ({\cal A}_{\textrm{add}}(R'))'\,,  \label{causality}
\ee
where $R'$ is the set of points spatially separated from $R$, and ${\cal A}'$ is the algebra of operators that commute with those in ${\cal A}$. If the inclusion (\ref{causality}) is not saturated for certain $R$, there is a larger algebra associated with $R$ ---the ``maximal algebra'', ${\cal A}_{\textrm{max}}(R)\equiv({\cal A}_{\textrm{add}}(R'))'$--- 
which necessarily contains a 
set $\{a\}$ of non-locally generated operators in region $R$ such that
\be 
{\cal A}_{\textrm{max}}(R)={\cal A}_{\textrm{add}}(R)\vee \{a\}\;.
\ee
For example, in free Maxwell theory, magnetic fluxes over open surfaces are, in this sense, non-local operators associated with ring-like regions. Wilson loops are particular examples. Given this algebraic structure, one can define classes of operators $[a]$ by making the quotient of the maximal algebra by the additive algebra. A class is defined by a certain representative $a$ and all operators that arise from it by adding products of local operators in $R$.

A non-trivial conclusion follows. The inclusion (\ref{causality}) forces a ``dual'' inclusion in the complementary region $R'$. This follows from von Neumann's double commutant theorem ---see \cite{Casini:2020rgj,Casini:2021zgr}. One concludes that if there exist non-local operators $\{a\}$ associated with $R$, there must be non-local operators $\{b\}$ associated with the complementary region $R'$ 
 \be 
{\cal A}_{\textrm{max}}(R')={\cal A}_{\textrm{add}}(R')\vee \{b\}\;.
\ee
Hence, non-local operator algebras come in dual pairs ---see appendix \ref{genMax} for a novel realization of this principle for generalized Maxwell fields. In fact, the ``size'' of these dual algebras is, in a precise sense, always the same. This is measured by the so-called ``Jones index'' 
\cite{Jones1983,KOSAKI1986123,longo1989} ---see also \cite{teruya,giorlongo,Magan:2020ake} for simpler introductions and specific computations in this context.

There are two further consequences from this approach. Consider a QFT with an additive algebra 
charged under a global symmetry group $G$. The question is whether this action can change the non-local classes of a given region $R$. The first consequence is that this can only happen in a point-like manner  \cite{Benedetti:2022zbb}, namely,
$
U(g)\, [a]\, U(g)^{-1} =[b]
$. 
If $[b]$ is different from $[a]$ for certain $[a]$ and certain $U(g)$, we say the classes are charged under the global symmetry. For continuous symmetry groups this implies that the non-local classes for $R$ must form a continuum. The second consequence is that if the non-local classes of $R$ are charged under the symmetry, the non-local classes for the complementary region $R'$ must necessarily be charged too  \cite{Benedetti:2022zbb}. Then they must also form a continuum of classes.

Going back to the discussion of conserved $p$-form currents, consider now integrating $\star\, J$ over an open surface, whose boundary $\partial \Sigma_{D-p}$ is a closed $(D-p-1)$-dimensional surface. The corresponding flux operator $\Phi$ ---defined as in (\ref{hfch1})---
has interesting topological properties. First, $\Phi$ only depends on the boundary $\partial \Sigma_{D-p}$. This implies that $\Phi$ commutes with all local operators outside $\partial \Sigma_{D-p}$. Second, since $\star\, J\,\neq \diff\,G$ with $G$ a physical field, $\Phi$ cannot be written as a circulation over the boundary of a physical field. If we consider a region $R$ enclosing the boundary $\partial\Sigma_{D-p}$ and with the same topology, this region will contain operators ---namely, $\Phi$ and all operators arising from multiplying it with local operators in $R$--- that commute with all local operators in $R'$, but cannot be locally generated in $R$, showing that
$
{\cal A}_{\textrm{add}}(R) \subsetneq  {\cal A}_{\textrm{max}}(R)
$.

 As a consequence of our previous discussion, an analogous strict inclusion takes place in $R'$.
Moreover, if the original classes of $R$ are charged under a continuum symmetry group, the dual classes in $R'$ form also a continuum, and are generated by a dual conserved $(D-p)$-form current $\tilde{J}$ (assuming a generalized version of Noether´s theorem). 
Hence, in certain scenarios, such as the one of linearized gravities, the existence of a conserved $p$-form current predicts the existence of a dual conserved $(D-p)$-form current.

{\bf Generalized Symmetries for Linearized Einstein gravity:}
A small perturbation $h_{\mu\nu}$ on top of Minkowski spacetime is defined by
\begin{equation}
   g_{\mu\nu}=\eta_{\mu\nu}+h_{\mu\nu}\, , \quad ||h_{\mu\nu}||\ll 1,\, \quad h_{\mu\nu}=h_{\nu\mu} \, .
\end{equation}
From the above expression for the metric  we can compute the associated quantities to the desired order in $h_{\mu\nu}$. 
In particular, the Einstein-Hilbert action reduces, at quadratic order in the perturbation, to the Fierz-Pauli one, 
\bea
 S_{\rm \ssc FP}= 
\int \diff^Dx\, \hspace{-0.35cm} && \left[\ \frac{1}{2}\partial_\lambda h^{\mu\nu} \partial_\nu h_{\mu\lambda}-\frac{1}{2}\partial_\mu h \partial_\nu h^{\mu\nu}\right. \label{fp}\\\notag && \,\, \left.+\frac{1}{4}\partial_\mu h\partial^\mu h-\frac{1}{4}\partial^\lambda h^{\mu\nu} \partial_\lambda h_{\mu\nu}\right]\, .
\eea
This action has a gauge-like symmetry under linearized diffeomorphisms acting as 
\be  \label{lineadif}
h_{\mu\nu} \to h_{\mu\nu} + \partial_\mu \xi_{\nu} + \partial_\nu \xi_{\mu}\, ,
\ee
and its variation with respect to $h_{\mu\nu}$ yields the linearized Einstein equations, 
\begin{equation}
    R_{\mu\nu}^{(1)}=0\, . \label{RR}
\end{equation}
The $p$-form conserved currents must be physical operators, {\it i.e.,} gauge-invariant with respect to transformations of the form (\ref{lineadif}). This means they must be written in terms of the linearized Riemann tensor $R_{\mu\nu\rho\sigma}$ ---or, equivalently, the Weyl--- which is the generator of the physical local algebra of the theory \cite{longo2019split,casini2021generalized}. 
Naturally, we are also free to use $\eta_{\mu\nu}$ as well as the Levi-Civita symbol $\varepsilon_{\mu_1...\mu_{D}}$ to build the currents. In fact, it turns out to be convenient to use the dual of the Riemann tensor, which is defined \cite{Hull:2001iu,Henneaux:2019zod} as
\be
 R^*_{\mu_1...\mu_{D-2}\alpha\beta}\equiv \frac{1}{2}\,\varepsilon_{\mu_1...\mu_{D-2}\lambda\sigma} \,R^{\lambda\sigma}_{\,\,\,\,\,\,\,\,\,\,\alpha\beta}\, .
\ee
A list of useful on-shell properties satisfied by the Riemann tensor and its dual appear in appendix \ref{appLin}.

{\it Four dimensions.}
Let us consider the $D=4$ case first. One finds the following four families 
of conserved two-forms \cite{casini2021generalized,Benedetti:2022zbb,hofman}
\bea\label{fam1}
{\sf A}_{\mu\nu} &\equiv & R^{(1)}_{\mu\nu\alpha\beta} \, a^{\alpha\beta}\,, \quad \label{A}\\
{\sf B}_{\mu\nu} &\equiv& R^{(1)}_{\mu\nu\alpha\beta}\,(x^\alpha b^\beta-x^\beta b^\alpha)\,, \quad\label{B}\\
{\sf C}_{\mu\nu} &\equiv& R^{(1)}_{\mu\nu\alpha\beta}\,c^{\alpha\beta\gamma}x_\gamma\,,\quad \label{C}\\
{\sf D}_{\mu\nu} &\equiv& R^{(1)}_{\mu\nu\alpha\beta}\,(x^\alpha d^{\beta\gamma}x^\gamma-x^\beta d^{\alpha\gamma}x^\gamma+\frac{1}{2} d^{\alpha\beta} x^2)\,.\quad \label{D}
\eea 
Here, $a^{\alpha\beta},\, b^\alpha,\, c^{\alpha\beta\gamma}$ and $d^{\alpha\beta}$  are real skew-symmetric free  parameters  which label each of the $20$ independent two-forms \footnote{Although charged under spacetime symmetries, these labels do not transform as Lorentz tensors \cite{Benedetti:2022zbb}.}. The charges satisfy the appropriate conservation equations
\be
\diff \,\star {\sf A}=\diff \,\star {\sf B}=\diff \,\star {\sf C}=\diff \,\star {\sf D}=0\;.
\ee
The conservation of ${\sf A}$ uses the fact that the Riemann tensor is itself a conserved current on shell. The conservation of ${\sf B}$ relies on the Einstein equation. For ${\sf C}$, the first Bianchi identity is used. Finally, for ${\sf D}$ we need both the Einstein equation and the first Bianchi identity.

Similar conserved currents $\{\tilde {\sf A},\tilde {\sf B},\tilde {\sf C},\tilde {\sf D}\}$ can be constructed from the dual curvature, by simply replacing $R^{(1)}_{\mu\nu\rho\sigma}$ by $R^{(1)*}_{\mu\nu\rho\sigma}$ and $\{a^{\alpha\beta},\, b^\alpha,\, c^{\alpha\beta\gamma},d^{\alpha\beta}\}$ by a new set of constant arrays
$\{\tilde a^{\alpha\beta},\, \tilde b^\alpha,\, \tilde c^{\alpha\beta\gamma},\tilde d^{\alpha\beta}\}$ in the definitions  (\ref{A})-(\ref{D}). Their conservation is more or less straightforward to verify.
However, observe  that in $D=4$ we have a $U(1)$ duality symmetry which corresponds to a rotation of the Riemann and its dual,
\bea 
\begin{pmatrix} R \\ R^\ast  \end{pmatrix} \to \begin{pmatrix} \cos \theta& -\sin\theta \\ \sin\theta & \cos\theta \end{pmatrix}\begin{pmatrix} R \\ R^\ast  \end{pmatrix}\,,
\eea
analogous to the one of the free Maxwell field. This means that the algebra generated by $\{{\sf A,B,C,D}\}$ is in fact the same as the one generated by $\{\tilde{{\sf A}},\tilde{{\sf B}},\tilde{{\sf C}},\tilde{{\sf D}}\}$.

In four dimensions, the fact that symmetries come in dual pairs might appear trivial at first sight since for a ring-like region the complement is also a ring. 
Still, it is not a coincidence that the total number of them is even ---namely, $20$--- and they can be seen to be organized in dual pairs when computing commutators \cite{Benedetti:2022zbb}. Notice that the currents should come in dual pairs since the conserved $2$-form currents (\ref{fam1}) are charged under spacetime symmetries. 

{\it General dimensions.}
We now move to $D>4$ dimensions. In that case, the complementary of a ring-like region is no longer a ring, so charges and dual charges will correspond to regions with different topologies. At first sight, following \cite{hofman}, one notices that the $2$-form families ${\sf A,B,C,D}$ described above are still conserved in general dimensions, which gives rise to $D(D+1)(D+2)/6$ candidates to generate generalized symmetries associated with rings. The principle that generalized symmetries come in dual pairs then predicts we should find an equal number of dual conserved $(D-2)$-forms.  Natural candidates appear by considering the obvious extension of the families $\tilde{{\sf A}},\tilde{ {\sf B}},\tilde{{\sf C}},\tilde{ {\sf D}}$ to higher dimensions. However, for $D> 4$ we  only recover $D(D+1)/2$ conserved $(D-2)$-forms in this way. These  are the two families $\tilde{ {\sf A}},\tilde{ {\sf B}}$ constructed as
\bea\label{tildfam}
\tilde{{\sf A}}_{\mu_1 \mu_2...\mu_{D-2}} &\equiv & R^{(1 )*}_{\mu_1 \mu_2...\mu_{D-2}\alpha\beta} \, \tilde{a}^{\alpha\beta}\,,\label{AT}\\
\tilde{{\sf B}}_{\mu_1 \mu_2...\mu_{D-2}} &\equiv & R^{(1 )*}_{\mu_1 \mu_2...\mu_{D-2}\alpha\beta}\,(x^\alpha \tilde{b}^\beta-x^\beta \tilde{b}^\alpha)\,.\label{BT}
\eea
The problem is that in $D>4$ we cannot build conserved currents of the $\tilde {\sf C}_{\mu\nu}$ and $\tilde {\sf D}_{\mu\nu}$ type. This is because the Bianchi identity of the dual Riemann tensor with only three indices contracted does not hold
\be \label{cerono}
\varepsilon^{\mu_1...\mu_{D-3}\alpha\beta\gamma}\, R^{(1 )*}_{\nu_1 ...\nu_{D-3}\alpha\beta\gamma}=\frac{1}{2}\eta_{\alpha\delta\nu_1 ...\nu_{D-3}}^{\beta\gamma\mu_1...\mu_{D-3}} R^{(1)\delta\alpha }_{\,\,\,\,\, \beta\gamma} \,. 
\ee
 While in $D=4$ this reduces to a combination of Ricci tensors which vanishes by virtue of the Einstein equation, this is no longer the case in $D>4$. 

This mismatch between the number of conserved charges associated with generalized symmetries in complementary regions has two possible origins. The first is that we might be missing charges arising from new conserved $(D-2)$-forms. In this case, the charges have to be of the $\tilde {\sf C}$ and $\tilde {\sf D}$ types, because potential  non-vanishing commutators with the ${\sf A}$'s and ${\sf B}$'s should be dimensionless (a c-number) ---see below. We argue these charges do not exist in appendix \ref{AppCBtildes}. The second possibility is that some of the conserved ${\sf A,B,C,D}$ two-form currents become exact in $D>4$ and do not generate generalized symmetries. Although counter-intuitive at first sight, this turns out to be the case. 
In order to see this, define
\bea \label{Atri}
\mathcal{A}_{\mu\nu\rho}&\equiv &- \frac{R^*_{\mu\nu\rho\alpha_1 ...\alpha_{D-3}}}{(D-4)!}\,\tilde{a}^{\alpha_1 ...\alpha_{D-3}\sigma}\,x_\sigma\,,\\ \label{exact}
\mathcal{C}_{\mu\nu\rho}&\equiv & \frac{R^*_{\mu\nu\rho\alpha_1 ...\alpha_{D-3}}}{(D-5)!(D-2)}\,\left(\,\frac{1}{2}\,\tilde{c}^{\alpha_1 ...\alpha_{D-3}}\,x^2 \right. \\ \notag &&\left.+\, \frac{\eta^{\alpha_1 ...\alpha_{D-3}}_{\beta_1 ...\beta_{D-3}}}{(D-4)!}\, c^{\beta_1 ...\beta_{D-4}\sigma} \,x^{\beta_{D-3} }\, x_\sigma\right)\,, 
\eea
where
$
\tilde{a}^{\alpha_1 ...\alpha_{D-2}}\equiv \frac{1}{2}\,\epsilon^{\alpha_1 ...\alpha_{D-2}\mu\nu}\,a_{\mu\nu}$ and  $\tilde{c}^{\alpha_1 ...\alpha_{D-3}}\equiv  \frac{1}{3!}\,\epsilon^{\alpha_1 ...\alpha_{D-3}\mu\nu\rho}\,c_{\mu\nu\rho}
$.
By direct computation, we find the corresponding divergences to be given by ${\sf A}$ and $ {\sf C}$ respectively, \ie
$
\partial^\rho \mathcal{A}_{\mu\nu\rho} =  {\sf A}_{\mu\nu}$ and $\partial^\rho \mathcal{C}_{\mu\nu\rho} =  {\sf C}_{\mu\nu}$. In differential-form language, we have
\bea
\diff \,\star \, \mathcal{A} \propto \, \star\, {\sf A}\,, \qquad \diff \,\star \,\mathcal{C} \propto \,\star \,{\sf C} \;.
\eea
Hence, $\star\, {\sf A}$ and $\star\, {\sf C}$ are exact in the physical algebra of the theory in $D>4$. Fluxes constructed from them belong to the additive algebra of the ring and, therefore, do not generate new generalized symmetries. Note that (\ref{Atri}) and (\ref{exact}) do not correspond to skew-symmetric differential forms in $D=4$. One could try to antisymmetrize the free indices but such a procedure results in both $\mathcal{A}_{\mu\nu\rho}$ and $\mathcal{C}_{\mu\nu\rho}$  vanishing identically. 

Summarizing, for $D>4$ we find two families of conserved $2$-forms generating generalized symmetries: the ${\sf B}$'s and the ${\sf D}$'s ---see (\ref{B}) and (\ref{D}). They generate a total of $D(D+1)/2$ conserved charges. In the complementary regions we also find two families of conserved $(D-2)$-forms generating generalized symmetries. These are the $\tilde{ {\sf A}}$'s and the $\tilde{{\sf B}}$'s ---see (\ref{AT}) and  (\ref{BT}). They generate an equal number of $D(D+1)/2$ conserved charges. The two sets contain the same number of charges and also have the right dimensions to produce non-vanishing quantum commutators. Starting from the ADM formalism \cite{1959PhRv..116.1322A}, we have evaluated such commutators explicitly in appendix \ref{commu}, finding the most-general non-vanishing ones to read  
\begin{align}
 \begin{split}
    &\left[\int_{\Sigma_{(D-2)}} \left(\star\, {\sf B}+\star\, {\sf D}\right), \int_{\Sigma_{2}} \left(\star\, {\tilde{\sf B}}+\star\, {\tilde{\sf A}}\right)\right]\\  & ={\rm i}(D-3) \left(2 \eta_{\mu\nu} b^{\mu} \tilde b^{\nu}+ \eta_{\mu\nu}\eta_{\rho\sigma}d^{\mu\rho}\tilde a^{\nu\sigma}\right)\, ,
     \end{split}
\end{align}
for regions whose boundaries have non-trivial linkings.
This verifies that the ${\sf B}$'s are paired with the $\tilde{{\sf B}}$'s and that the ${\sf D}$'s are paired with the $\tilde{{\sf A}}$'s, in agreement with the four-dimensional results of \cite{Benedetti:2022zbb}.


An interesting question is if in $D\geq 6$ one can find  conserved $p$-forms with $p\neq 2$ and $p\neq (D-2)$. This is the case, but in all cases we found that such conserved currents are always exact forms and therefore do not generate new symmetries.


{\bf Generalized Symmetries for Linearized Higher-Curvature Gravities:}
Let us consider now a 
general gravity action built from contractions of the Riemann tensor and the metric
\begin{equation}\label{LR2}
S=\frac{1}{16\pi G}\int \diff ^Dx\,\, \sqrt{|g|}\, \mathcal{L}(g^{\alpha\beta},R^{\rho}_{\,\,\, \sigma \mu\nu})\, .
\end{equation}
As explained in appendix \ref{hogs}, the most general theory of this form contributing non-trivially to the linearized equations of motion on Minkowski space is quadratic in the Riemann tensor, namely,
\begin{equation}\label{quadi}
    \mathcal{L}= R + \alpha_1 R^2 + \alpha_2 R_{\mu\nu}R^{\mu\nu}+\alpha_3 R_{\mu\nu\lambda\sigma}R^{\mu\nu\lambda\sigma}\,,
\end{equation}
where $\alpha_{1,2,3}$ are arbitrary constants with dimensions of length squared. 
Following \cite{Bueno:2016ypa}, it is convenient to write the couplings $\alpha_1$ and $\alpha_2$ in terms of two new parameters, $m_s$ and $m_g$, as 
\be
\alpha_1\equiv \frac{(D-2)m_g^2+ D m_s^2 }{4(D-1) m_s^2\, m_g^2}+\alpha_3\, ,\quad \alpha_2\equiv   \frac{-1}{m_g^2} -4\alpha_3\, .
\ee 
The linearized equation resulting from the theory can then be written as
\begin{equation}\label{eeom}
    \mathcal{E}_{\mu\nu}^{(1)}\equiv \left[1-\frac{\partial^2}{m_g^2}\right]R^{(1)}_{\mu\nu}-\Delta_{\mu\nu}R^{(1)}=0\, ,
\end{equation}
where we defined the operator
\be
\Delta_{\mu\nu}\equiv \frac{\eta_{\mu\nu}}{2} \left[1-\frac{\partial^2}{m_g^2} \right]+\frac{(D-2)(m_g^2-m_s^2)}{2(D-1)m_s^2 m_g^2} \left[\partial_{\mu} \partial_{\nu}-\eta_{\mu\nu} \partial^2\right].
\ee
Eq.\,(\ref{eeom}) 
reduces to the linearized Einstein one for $m_g^2,m_s^2\rightarrow \infty$, which is equivalent to turning off  the quadratic couplings. As explained \eg in \cite{Bueno:2016ypa} ---see appendix \ref{hogs} for a brief review--- $m_g^2$ and $m_s^2$ correspond to the squared-masses of two ---spin-0 and spin-2, respectively--- extra modes which appear in the linearized spectrum of these theories 

Since the gauge symmetry (\ref{lineadif}) remains the same, the Riemann tensor is still the generator of gauge invariant operators. 
However, the Ricci tensor does not vanish on shell anymore and, consequently, $R^{(1)}_{\mu\nu\rho\sigma}$ is no longer a conserved current. 
On the other hand, the properties of the dual Riemann tensor only depend on its symmetries and the Bianchi identities. Hence, we obtain the very same set of
$D(D+1)/2$ independent $(D-2)$-form conserved currents corresponding to the families $\tilde{ {\sf A}}$ and $\tilde{ {\sf B}}$ as for Einstein gravity ---see (\ref{tildfam}) and (\ref{BT}). 

Since these currents are again charged under spacetime symmetries, this result should imply the existence of an equal number of $D(D+1)/2$ dual $2$-form conserved currents. 
Equivalently, the dual-pairs principle suggests the existence of a generalized tensor playing the role of the Riemann. 
A candidate is given by
\begin{align}\notag
\mathcal{W}_{\mu\nu\alpha\beta} \equiv & \,\mathcal{R}_{\mu\nu\alpha\beta} + \frac{2}{(D-2)} \,\Big[ \eta_{\nu[\alpha} \mathcal{R}_{\beta]\mu}-\eta_{\mu[\alpha} \mathcal{R}_{\beta]\nu} \Big] \\ &+  \frac{2}{(D-2)(D-1)}  \eta_{\mu [\alpha }\eta_{\beta]\nu} \, \mathcal{R}\, , \label{Jw}
\end{align}
where we defined 
\be
\mathcal{R}_{\mu\nu\alpha\beta}\equiv \left[1-\frac{\partial^2}{m_g^2} \right]R^{(1)}_{\mu\nu\alpha\beta}+2\Delta_{\mu[\beta}R^{(1)}_{\alpha]\nu}+2\Delta_{\nu[\alpha|}R^{(1)}_{\beta]\mu}\, . \label{JJ}
\ee
The tensor $\mathcal{W}_{\mu\nu\alpha\beta}$ is traceless  and satisfies its own Bianchi identity, $\eta^{\mu\alpha}\mathcal{W}_{\mu\nu\alpha\beta} = 0$, $  \varepsilon^{\mu_1...\mu_{D-3}\alpha\beta\gamma}\,\mathcal{W}_{\alpha\beta\gamma\nu}=0 $. On the other hand, one can show that it is a conserved current on-shell and that the second Bianchi identity holds only in two cases: i)  when the spin-2 mode is absent from the spectrum ($m_g^2\rightarrow \infty$); ii) when $m_s^2=m_g^2$. In those situations, we find that
$
    \partial^{\mu} \mathcal{W}_{\mu\nu\alpha\beta} =0 
$
and $ \varepsilon^{\mu_1...\mu_{D-3}\alpha\beta\gamma}\,\partial_\alpha\, \mathcal{W}_{\beta\gamma\mu\nu}=0$. 
In the first case, the quadratic part of the action reduces to a single $R^2$ term, whereas in the second, we are left with a single Weyl$^2$ term. 
In both situations, it follows that $\mathcal{W}_{\mu\nu\alpha\beta}$ generates non-trivial charges of the ${\sf B}$ and ${\sf D}$ classes identical to the Einstein gravity ones ---see (\ref{B}) and (\ref{D})--- by  simply replacing $R_{\mu\nu\rho\sigma}^{(1)}$ with $\mathcal{W}_{\mu\nu\rho\sigma}$. Analogously, one can show that the putative ${\sf A}$ and ${\sf C}$ ones are exact in  a similar way to (\ref{Atri}) and (\ref{exact}).

One may be tempted to think that $\mathcal{R}_{\mu\nu\alpha\beta}$ could define additional non-trivial charges. This is not the case though. Similarly, one can define ${\sf \tilde A}$ and ${\sf \tilde B}$ charges using the dual of $\mathcal{W}_{\mu\nu\alpha\beta}$ instead of the dual of the Riemann tensor. However, one can show that such charges produce the same non-local classes as the ones defined in (\ref{tildfam}) and (\ref{BT}). These claims are  proven in appendix \ref{GSHG}.

Summarizing, in the absence of the spin-2 mode we find that the higher-curvature theories possess  $D(D+1)$ conserved currents, organized in two equal-size dual sets $\{ \tilde A, \tilde B\}$ and $\{ B, D \}$. The currents are continuous deformations of the Einsteinian ones, obtained by replacing $R_{\mu\nu\alpha\beta}^{(1)}$ by $\mathcal{W}_{\mu\nu\alpha\beta}$ in the corresponding expressions. 

When $m_g^2$ is finite and generic, this construction fails and we find a violation of the dual-pairs principle. This is likely related to the fact that the spin-2 mode is a ghost \cite{Bueno:2016ypa,Alvarez-Gaume:2015rwa}, whose presence renders the theory nonunitary \cite{Stelle:1976gc,Stelle:1977ry}. It is reasonable to expect generalized symmetries and the dual-pairs principle to be sensitive to such issue. Nonetheless, it is also a logical possibility that a more elusive set of charges exists in this case and ends up saving the day for these theories.  

On the other hand, the case $m_g^2=m_s^2$ 
has similar unitarity problems \cite{Stelle:1977ry}. The fact that this does not violate the dual-pairs principle suggests that consistent theories will always respect such principle, but that the opposite implication will not be true in general.

{\bf Conclusions and future work:}
In this letter we have found $D(D+1)$ generalized symmetries for linearized Einstein gravity as well as for higher-curvature gravities propagating an additional spin-0 mode in general dimensions. Half of the symmetries are generated by 
 $2$-form currents and the other half by $(D-2)$-form currents, which verifies the QFT principle that generalized symmetries always come in dual pairs. 
In the case of higher-curvature gravities propagating an additional massive spin-2 graviton, the theory is nonunitarity, and the dual-pairs principle seems to be violated. 
 

An interesting outcome of our analysis is that generalized gravitons can be defined 
by their generalized symmetries, supporting the perspective developed in \cite{hofman} ---see appendix \ref{gravfromsym}. More precisely, linearized gravity is a theory of symmetry, fully characterized by the conservation of its closed-form currents. This parallels the case of the Maxwell field. It also reminds to the non-perturbative completion given by AdS/CFT \cite{Maldacena:1997re,Aharony_2000}. In this scenario, gravity is dual to the dynamics of the dual CFT stress tensor, which is fully constrained by its 
conservation, 
its tracelessness and the associated Ward identities.

Further interesting outcomes are that 
the graviton generalized symmetries are charged under spacetime symmetries. Following \cite{Benedetti:2022zbb} this result implies the Weinberg-Witten theorem \cite{WEINBERG198059} for these theories. Also, the same result shows that these theories enlarge the space of so-called tensor gauge theories \cite{Rasmussen2016}. 
In particular, they provide further examples of the recently proposed connection between gravity and fractonic systems \cite{Pretko:20161,Pretko:20162,Pretko:2017,Benedetti:2022zbb,hofman,Blasi:2022mbl,Bertolini:2023juh}.

There are several venues for future work. First, our analysis could be extended to: linearized theories of gravity whose Lagrangian is also a functional of the covariant derivative; theories with explicit mass terms in the action
; more general backgrounds. 
A somewhat more difficult question is whether the existence of a non-trivial space of low-energy gravity theories ---defined in terms of the spectrum of generalized symmetries--- implies the existence of a similar space of UV completions. One expects this not to be the case, and that the absence of generalized symmetries in quantum gravity \cite{Polchinski:2003bq,Banks:2010zn} should lead to a unified theory in the UV, where all these different phases are smoothly connected to each other. 

{\bf Acknowledgements.}  We would like to thank Pablo A. Cano, Horacio Casini, Tom\'as Ort\'in and specially Diego Hofman for useful discussions.
PB was supported by a Ram\'on y Cajal fellowship (RYC2020-028756-I) from Spain's Ministry of Science and Innovation. The work of VB and JM is supported by CONICET, Argentina.

\onecolumngrid  \vspace{0.2cm} 
\appendix 

\section{Generalized symmetries for generalized Maxwell fields}\label{genMax}

In this appendix we study the generalized symmetries of a $U(1)$ gauge field in $D$ spacetime dimensions. We consider theories with higher-order interaction terms that only depend on the field strength. These are called ``Non-linear electrodynamics'' (NLE) theories ---see \cite{Russo:2022qvz, Guerrieri:2022sod,Benedetti:2022ofj,Cano:2021tfs,Denisov:2017qou} for recent literature on the subject and \cite{Sorokin:2021tge} for a recent review and references. Interesting examples within this class of theories include Born-Infeld \cite{Born:1934gh} and the more recent ``ModMax'' \cite{Bandos:2020jsw,Bandos:2021rqy,Kosyakov:2020wxv} and Quasi-topological electromagnetism theories \cite{Liu:2019rib,Cisterna:2020rkc}. Except for the free Maxwell theory, these theories are non-renormalizable. At the quantum level, they can only be considered as effective field theories below certain scale. Indeed, Ref.\,\cite{Benedetti:2022ofj} shows that these theories can only be defined at all energy scales by introducing charged fields at certain energy, explicitly breaking the generalized symmetries. Here we are only discussing them as effective field theories.  To be specific, NLE theories are described by the Lagrangian density
 \be 
 \mathcal{L}=-\frac{1}{4} F_{\mu\nu}F^{\mu\nu}\,+\,\mathcal{L}_I(F_{\mu\nu})\,,\quad F_{\mu\nu}= \partial_\mu A_\nu- \partial_\nu A_\mu\,,
 \label{vlag}
 \ee
 where $\mathcal{L}_I(F_{\mu\nu})$ is a generic Lorentz invariant functional of the field strength. Observe that we have not included an explicit dependence of the field $A_\mu$, but only of its derivatives. In this zoo of theories, we can find two dual conserved currents. First, from $F=\diff\,A$
 we have that
 \be 
 \tilde{F}^{\mu_1\mu_2\,...\, \mu_{D-2}}\equiv \frac{1}{2}\varepsilon^{\mu_1\mu_2\,...\, \mu_{D-2} \rho\sigma }F_{\rho\sigma}\;,
 \ee
 is a conserved $(D-2)$-form current since it obeys $\diff\,\star\, \tilde{F}=0$. Second, the equations of motion can be expressed as 
 \be 
 \partial_\nu\left( \frac{\partial \mathcal{L}}{\partial (\partial_{\nu} A_{\mu})}\right)=   \partial_\nu \left( F^{\mu \nu} + \frac{\partial \mathcal{L}_I}{\partial (\partial_{\nu} A_{\mu})}\right)=   \partial_\nu \left( F^{\mu \nu} + \frac{\partial \mathcal{L}_I}{\partial  F_{\nu \mu}}- \frac{\partial \mathcal{L}_I}{\partial  F_{\mu \nu}}\right)=0\,,
 \label{vmov}
 \ee
and allow us to define the $2$-form gauge invariant current 
 \be 
 \mathcal{F}= \frac{1}{2} \mathcal{F}_{\mu\nu}\, \diff x^\mu\wedge \diff x^\nu \,, \,\quad  \mathcal{F}_{\mu\nu} \equiv  F^{\mu \nu} + \left( \frac{\partial \mathcal{L}_I}{\partial  F_{\nu \mu}}- \frac{\partial \mathcal{L}_I}{\partial  F_{\mu \nu}}\right)\;,
 \label{vcur1}
 \ee
 which obeys the conservation law $\diff\,\star\, \mathcal{F}=0$. 
 
 The conserved charges associated with these $(D-2)$-form and $2$-form currents can be obtained by integrating their Hodge duals over $2$ and $(D-2)$ dimensional surfaces respectively. These expressions can be used to compute the commutation relations between the dual charges. Using canonical quantization techniques, one obtains that this commutator is proportional to the linking number between the surfaces defining it, generalizing the free Maxwell field result. This computation can be done by introducing the smeared version of the non-local operators, a method developed recently in \cite{Pedro,casini2021generalized} or in the same manner as performed for Einstein gravity in  appendix \ref{commu}. Specifically, the result yields for surfaces whose boundaries are only linked once 
 \be
    \left[\int_{\Sigma_{(D-2)}} \star \mathcal{F}, \int_{\Sigma_{2}} \star\, {\tilde{ F}}\right]={\rm i}\,.
\ee
 Note that in $D=4$ both conserved currents $\mathcal{F}$ and $\tilde{F}$ are $2$-forms, generating charges with support in $2$-dimensional surfaces. This implies that both of them can produce non-local operators in the same region  and the complement  of such a region.  This is not true in $D\neq 4$ where the charges generated by $\star \mathcal{F}$ have to be assigned to $(D-2)$-dimensional surfaces  with $(D-2)\neq 2$. 

 \section{On-shell properties of the Riemann tensor and its dual}\label{appLin}
In this appendix we list a set of properties satisfied by the Riemann tensor and its dual for spacetimes fulfilling the Einstein equations.  For the Riemann tensor we have
\begin{alignat}{3} \label{SS}
& R_{\mu\nu\alpha\beta}= -R_{\nu\mu\alpha\beta}=-R_{\mu\nu\beta\alpha}\,,\quad &&\text{[Antisymmetry]}\\ 
& R_{\mu\nu\alpha\beta}=R_{\alpha\beta\mu\nu}\,,\quad &&\text{[Interchange]} \label{IS}\\
& \eta^{\mu\alpha}\, R_{\mu\nu\alpha\beta}=0\,,\quad && \text{[Einstein]}\label{EE} \\
& \varepsilon^{\mu_1...\mu_{D-3}\alpha\beta\gamma}\, R_{\alpha\beta\gamma\nu}=0\,,\quad && \text{[1st Bianchi]}\label{FB}
 \\ & \varepsilon^{\mu_1...\mu_{D-3}\alpha\beta\gamma}\,\partial_\alpha\, R_{\beta\gamma\mu\nu}=0\,,\quad &&\text{[2nd Bianchi]}\label{sbi}
  \\ & \partial^\mu \,R_{\mu\nu\alpha\beta}=0 \,.\quad&& \text{[Einstein and 2nd Bianchi}]
  \label{CR7}
\end{alignat}
On the other hand, the dual Riemann tensor satisfies the following algebraic and conservation equations
 \begin{alignat}{3}
&R^*_{\mu_1 \mu_2...\mu_{D-2}\alpha\beta}= -R^*_{\mu_2 \mu_1...\mu_{D-2}\alpha\beta}=...\,, \quad &&  \text{[Levi-Civita antisymemtry]}  \label{LSS}\\
&R^*_{\mu_1 \mu_2...\mu_{D-2}\alpha\beta}=-R^*_{\mu_1 \mu_2...\mu_{D-2}\beta\alpha}\,,\quad && \text{[Riemann antisymmetry]}\label{RSS}\\
& \eta^{\gamma\alpha}\, R^*_{\gamma \mu_1...\mu_{D-3}\alpha\beta}=0\,,\quad &&\text{[1st Bianchi]}\label{TDR}\\
&\varepsilon^{\mu_1 \mu_2...\mu_{D-1}\beta}R^*_{\mu_1 \mu_2...\mu_{D-1}\alpha}= 0\,, \quad &&\text{[Einstein]}\label{ED}\\
&\varepsilon^{\mu_1 \mu_2...\mu_{D-1}\beta}R^*_{\alpha\mu_1 \mu_2...\mu_{D-1}}= 0\,, \quad &&\text{[Einstein]} \label{ED2}\\
& \partial^\gamma \,R^*_{\gamma \mu_1...\mu_{D-3}\alpha\beta}=0\,, \quad &&\text{[2nd Bianchi]} \label{CDR} \\
& \partial^\beta \,R^*_{\mu_1...\mu_{D-2}\alpha\beta}=0\,, \quad &&\text{[Riemman conservation]} \label{CDR2} \\
& \varepsilon^{\mu_1 \mu_2...\mu_{D-1}\gamma}\partial_{\mu_1}R^*_{\mu_2 \mu_3...\mu_{D-1}\alpha\beta}=0\,, \quad &&\text{[Riemman conservation]} \label{CDR3} \\
& \varepsilon^{\nu_1 \nu_2...\nu_{D-3}\alpha \beta \gamma}\partial_{\gamma}R^*_{\mu_1 \mu_2...\mu_{D-2}\alpha\beta}=0\, . \quad &&\text{[2nd Bianchi]} \label{CDR4}
\end{alignat}

\section{Looking for conserved $(D-2$)-forms in $D>4$ dimensions} \label{AppCBtildes}
When looking for conserved $(D-2)$-forms for the Einstein graviton in $D>4$ dimensions, the first proposal one might consider is the straight-up generalization of the $D=4$ case. This includes the  currents $\tilde{A}$ and $\tilde{B}$ defined in (\ref{AT}) and (\ref{BT}) as well as 
\bea
\tilde{C}_{\mu_1 \mu_2...\mu_{D-2}} &=& R^*_{\mu_1 \mu_2...\mu_{D-2}\alpha\beta} \,\tilde{c}^{\alpha\beta\gamma}\,x_\gamma,\label{CTp}\\
\tilde{D}_{\mu_1 \mu_2...\mu_{D-2}}  &=& R^*_{\mu_1 \mu_2...\mu_{D-2}\alpha\beta} \,\,\left(x^\alpha \tilde{d}^{\beta \gamma} x_\gamma + x^\beta \tilde{d}^{\gamma \alpha} x_\gamma + \frac{1}{2} \tilde{d}^{ \alpha\beta} x^2\right)\;,\label{DTp}
\eea 
where $\tilde{c}^{\alpha\beta\gamma}$ and $\tilde{d}^{\beta \gamma}$ are again real skew-symmetric tensors of free parameters. However, these are not conserved. Specifically their divergence yields
\bea
\partial^\rho \tilde{C}_{\mu_1 \mu_2...\mu_{D-3}\rho } &=& R^*_{\mu_1 \mu_2...\mu_{D-3}\rho \alpha\beta} \,\tilde{c}^{\rho \alpha\beta}\;,\label{CTdiv}\\
\partial^\rho \tilde{D}_{\mu_1 \mu_2...\mu_{D-3}\rho}  &=& R^*_{\mu_1 \mu_2...\mu_{D-3}\rho\alpha\beta} \,\,\left(x^\alpha \tilde{d}^{\beta \rho} + x^\beta \tilde{d}^{\rho \alpha} + \tilde{d}^{ \alpha\beta} x^\rho \right)\;.\label{DTdiv}
\eea 
Nevertheless, in $D>4$ the Bianchi identities of the dual Riemann tensor (\ref{ED}) and (\ref{ED2})  are less restrictive than in $D=4$ and allow for several modifications of this naive proposal. Indeed,  one might write all possible combinations of the dual Riemann tensor and its derivatives with $\tilde{c}^{\alpha\beta \gamma}$, $\tilde{d}^{\alpha \beta }$  and the spacetime coordinates that obey the required skew-symmetry and have the right scaling dimension. Many of these combinations are not linearly independent of each other, but, making various algebraic manipulations that involve adding certain terms to the naive forms of the families $\tilde C$ and $\tilde D$, one finds additional conserved currents. More precisely, defining
\bea
\tilde{C}_{\mu_1 \mu_2...\mu_{D-2}} &=& \big (R^*_{\mu_1 \mu_2...\mu_{D-2}\alpha\beta} + \frac{x^\sigma}{3} \partial_\sigma R^*_{\mu_1 \mu_2...\mu_{D-2}\alpha\beta}\big)\,\tilde{c}^{\alpha\beta\gamma}\,x_\gamma\nonumber \\ &-&\frac{1}{3 (D-3)!}\eta^{\nu_1 ... \nu_{D-2}}_{\mu_1 ... \mu_{D-2}} R^*_{\nu_1...\nu_{D-3}\alpha\beta\gamma}\, \tilde{c}^{\alpha\beta\gamma}\,x_{\nu_{D-2}} \,,\label{CTp2}\\
\tilde{D}_{\mu_1 \mu_2...\mu_{D-2}}  &=& \big (R^*_{\mu_1 \mu_2...\mu_{D-2}\alpha\beta} + \frac{x^\sigma}{4} \partial_\sigma R^*_{\mu_1 \mu_2...\mu_{D-2}\alpha\beta}\big)\,\,\left[x^\alpha \tilde{d}^{\beta \gamma} x_\gamma + x^\beta \tilde{d}^{\gamma \alpha} x_\gamma + \frac{1}{2} \tilde{d}^{ \alpha\beta} x^2\right]  \nonumber \\ &-&\frac{1}{4 (D-3)!}\eta^{\nu_1 ... \nu_{D-2}}_{\mu_1 ... \mu_{D-2}} R^*_{\nu_1...\nu_{D-3}\alpha\beta\gamma}\,x_{\nu_{D-2}} \, \left[x^\alpha \tilde{d}^{\beta \gamma} + x^\beta\tilde{d}^{\gamma \alpha}  + x_\gamma \tilde{d}^{ \alpha\beta}\right]\,,\label{DTp2}
\eea 
one gets the conservation laws 
\bea
\partial^\rho \tilde{C}_{\mu_1 \mu_2...\mu_{D-3}\rho } = 0\, , \qquad 
\partial^\rho \tilde{D}_{\mu_1 \mu_2...\mu_{D-3}\rho}  = 0\label{DTdiv2}\,.
\eea 
Although the number of these new current is the expected one, the problem is that these currents are exact and do not generate generalized symmetries. Namely, we can find $\tilde{\mathcal{C}} $ and $\tilde{\mathcal{D}} $ physical $(D-1)$-forms so that
\be
\diff \star \tilde{\mathcal{C}}  \sim \star\, {\tilde C}  \,,\qquad \diff \star \tilde{\mathcal{D}} \sim \star\, {\tilde D}\,.
\ee
The components of $\tilde{\mathcal{C}} $ and $\tilde{\mathcal{D}} $ are  given by
\bea
\tilde{\mathcal{C}}_{\mu_1 \mu_2...\mu_{D-1}}&=& \frac{\eta^{\nu_1 ... \nu_{D-1}}_{\mu_1 ... \mu_{D-1}}}{3 (D-2)!} R^*_{\nu_1...\nu_{D-2}\alpha\beta}\,c^{\alpha\beta \gamma}\,x_\gamma\,x_{\nu_{D-1}} \, \,, \\ 
\tilde{\mathcal{D}}_{\mu_1 \mu_2...\mu_{D-1}} &=& \frac{\eta^{\nu_1 ... \nu_{D-1}}_{\mu_1 ... \mu_{D-1}}}{4 (D-2)!} R^*_{\nu_1...\nu_{D-2}\alpha\beta}\,\left[x^\alpha \tilde{d}^{\beta \gamma} + x^\beta\tilde{d}^{\gamma \alpha}  + \frac{1}{2} x_\gamma \tilde{d}^{ \alpha\beta}\right]\,x_\gamma\,x_{\nu_{D-1}} \,.
\eea
Indeed, from these expressions, one can verify the claimed relations
\be
\partial^\rho \tilde{\mathcal{C}}_{\mu_1 \mu_2...\mu_{D-2}\rho} = {\tilde C}_{\mu_1 \mu_2...\mu_{D-2}} \,,\quad \partial^\rho \tilde{\mathcal{D}}_{\mu_1 \mu_2...\mu_{D-2}\rho} = {\tilde D}_{\mu_1 \mu_2...\mu_{D-2}}\,.
\ee

\section{Topological commutators for the Fierz-Pauli graviton }
\label{commu}
In this appendix, we evaluate the commutator between the fluxes defined as integrals of the  conserved $2$- and $(D-2)$-form currents for the Fierz-Pauli theory in $D\geq 4$. To begin, we notice that the Riemann tensor can be written to linear order in the perturbation field $h_{\mu\nu}$ as
\be 
R_{\mu\nu \alpha\beta} =\frac{1}{2}\left(\partial_\beta \partial_\mu h_{\nu\alpha} +\partial_\alpha \partial_\nu h_{\mu\beta} - \partial_\alpha \partial_\mu h_{\nu\beta} - \partial_\beta \partial_\nu h_{\mu\alpha}\right)\, .\label{rr}
\ee
However, for the purpose of this calculation,  it will be useful to re-write (\ref{rr})  as a function of the dynamical phase-space variables. We obtain that the components of the on-shell  Riemann tensor  read
\bea
& R_{0i0j} &=\frac{1}{2} \left(\partial_i \partial_n h_{nj}+\partial_j \partial_n h_{ni}-\partial_i \partial_j h_{nn} - \partial_n\partial_n h_{ij}   \right)\, ,\label{R00} \\
& R_{0ijk} &=\ \partial_l {\pi}_{ik}- \partial_k {\pi}_{il}-\frac{\delta_{ik}}{d-1}\,\partial_l\pi_{nn}+\frac{\delta_{il}}{d-1}\,\partial_k\pi_{nn}\,, \label{R0} \\
& R_{ijkl} &=\frac{1}{2}\left(\partial_l \partial_i h_{jk} +\partial_k \partial_j h_{il} - \partial_k \partial_i h_{jl} - \partial_l \partial_j h_{ik}\right)\,,\label{Rs}
\eea
where $\pi_{ij}$   represents the associated canonical momenta to $h_{ij}$, defined  from the Fierz-Pauli  lagrangian  (\ref{fp}) as
\be
\pi_{ij}\equiv \frac{\delta \mathcal{L}_{\rm \ssc FP}}{\delta \dot{h}^{ij}} = \frac{1}{2}(\dot{h}_{ij}-\partial_i h_{0j}-\partial_j h_{0i}-\delta_{ij} \dot{h}_{nn} +2 \delta_{ij} \partial_n h_{0n})\,.
\label{pi}
\ee

We now proceed by computing the commutator of the fluxes defined over a $(D-2)$-dimensional  finite spatial ``square'' $\Sigma_{D-2}$ and a $2$-dimensional one $\Sigma_2$. We choose the coordinate system as  $(x^0,\,x^1,x^2,...,x^{D-1} )$, so that the surfaces are defined by the length parameters $L,\,\alpha,\,$ and $\beta$ as
\begin{align}
&\Sigma_{D-2} \equiv \Big\{x^0=0\,, \,\,x^1=0,\,\,x^2 \in [0, \alpha ]\,, \,\, x^3,....,x^{D-1} \in [-L/2,L/2]\,\Big\}\,, \quad 0<\alpha<3L/2\,,\\
&\Sigma_2 \equiv \Big\{x^0=0\,, \,\,x^1 \in [-L/2,L/2]\,, \,\, x^2\in [\beta ,3L/2]\,, \,x^3,....,x^{D-1} =0\Big\} \,, \quad 0<\beta<3L/2\,.
\end{align}
Considering (\ref{R00}-\ref{Rs}) and the equal-time commutation relations  given by the canonical quantization  of the  Fierz-Pauli theory 
\be 
[h_{ij}(x),\pi_{kl}(y)]=\frac{ {\rm i}}{2}\big(\delta_{ik}\delta_{jl}+\delta_{il}\delta_{jk}\big)\delta(x-y) \, ,\label{CC}
\ee
or equivalently, the linearized ADM approach \cite{1959PhRv..116.1322A} ---see \cite{casini2021generalized} for a concise description--- we recover by direct computation the result
\be
    \left[\int_{\Sigma_{(D-2)}} \left(\star\, {\sf B}+\star\, {\sf D}\right), \int_{\Sigma_{2}} \left(\star\, {\tilde{\sf B}}+\star\, {\tilde{\sf A}}\right)\right]={\rm i}(D-3) \left(2 \eta_{\mu\nu} b^{\mu} \tilde b^{\nu}+ \eta_{\mu\nu}\eta_{\rho\sigma}d^{\mu\rho}\tilde a^{\nu\sigma}\right)\theta(\alpha-\beta)\, . \ 
\ee
The dependence of the result on the Heaviside function $\theta(\alpha-\beta)$ represents the fact that the commutator
is only non-vanishing when $\alpha>\beta$, namely,  when the boundaries of the squares are linked.  The conservation laws
\begin{equation}
\diff\star\, {\sf B}=\diff \star\, {\sf D}=\diff \star\, \tilde{\sf A}=\diff \star\, \tilde{\sf B}=0\, ,   
\end{equation}
imply that the same argument holds for any other pair of geometries $\Sigma_{D-2}$ and $\Sigma_{2}$. 
As long as their boundaries are linked, the commutator will be given in terms of $(b\cdot\tilde{b})$ and $(d\cdot\cdot\,\tilde{a})$ by the above expression. Whenever they are not, the commutator will vanish.

\section{Linearized higher-curvature gravities}\label{hogs}

In the absence of additional fields, the non-linear equations of motion of a general diffeomorphism-invariant theory of gravity built from contractions of the Riemann tensor and the metric and with Lagrangian density $ \mathcal{L}(g^{\alpha\beta},R^{\rho}_{\,\,\, \sigma \mu\nu})$,   read
\begin{equation}\label{eomG}
\mathcal{E}_{\mu\nu}\equiv P_{\mu}\,^{\sigma\rho\delta}R_{\nu\sigma\rho\lambda}-\frac{1}{2}g_{\mu\nu}\mathcal{L}-2\nabla^{\alpha}\nabla^{\beta}P_{\mu\alpha\beta\nu}=0\, , 
\end{equation}
where $ \quad P^{\mu}_{\,\,\, \nu\rho\sigma}\equiv \partial \mathcal{L}/ \partial R^{\mu}_{\,\,\,\nu\rho\sigma}$. 
Given a generic higher-curvature density of order $n$ in Riemann curvatures, observe that the three terms appearing in (\ref{eomG}) are of orders: Riem$^n$,  Riem$^n$ and Riem$^{n-1}$, respectively. Hence, at linear order in $h_{\mu\nu}$, each of those terms will be of the form: Riem$^{(1)}\cdot $ [Riem$^{n-1}$]$^{(0)}$, Riem$^{(1)}\cdot $ [Riem$^{n-1}$]$^{(0)}$ and Riem$^{(1)}\cdot $ [Riem$^{n-2}$]$^{(0)}$, respectively. Since the Riemann tensor  vanishes identically on the Minkowski background, all such terms will vanish for general $n$ with two single exceptions: i) the first two terms for $n=1$, which combined will be nothing but the linearized Einstein tensor; ii) the third term for $n=2$. Therefore, it suffices to consider a general quadratic theory to account for the most general case. This changes when a cosmological constant is present, since then Riem$^{(0)}\neq 0$ ---see \eg \cite{Sisman:2011gz,Lu:2011zk,Bueno:2016ypa,Bueno:2016xff}. 

We can write a general quadratic theory as \cite{Salvio:2018crh,Alvarez-Gaume:2015rwa}
\begin{equation}\label{quadi}
    \mathcal{L}=\frac{1}{16\pi G} \left[R + \alpha_1 R^2 + \alpha_2 R_{\mu\nu}R^{\mu\nu}+\alpha_3 R_{\mu\nu\lambda\sigma}R^{\mu\nu\lambda\sigma} \right]\,,
\end{equation}
where $\alpha_{1,2,3}$ are arbitrary constants with dimensions of length squared. Again, the linearized equations for this theory can be obtained from the action expanded to quadratic order in the perturbation. This can be written as 
\be
S=\frac{1}{16\pi G} \left[ S_{\rm \ssc FP} + S_{\rm \ssc HD}\right]+\mathcal{O}(h^3)\, ,
\label{FPHD}
\ee
where the first term, $S_{\rm \ssc FP}$, is the Fierz-Pauli action coming from the Einstein-Hilbert density  and reads
\bea 
S_{\rm \ssc FP} &=& \int \diff ^Dx\,\left[\left( 1+\frac{h}{2} \right) \, R^{(1)} + R^{(2)}\right] \label{FP}\\ &=& \int \diff^Dx\,\left[\ -\frac{1}{2}\partial_\mu h \partial_\nu h^{\mu\nu}+\frac{1}{2}\partial_\lambda h^{\mu\nu} \partial_\nu h_{\mu\lambda}+\frac{1}{4}\partial_\mu h\partial^\mu h-\frac{1}{4}\partial^\lambda h^{\mu\nu} \partial_\lambda h_{\mu\nu}\right]\, ,
\eea
and the second term, $S_{\rm \ssc HD}$, comes from the quadratic densities, yielding
\bea 
S_{\rm \ssc HD} &= & \int \diff ^Dx\,\left[\alpha_1\, R_{(1)}^2+ \alpha_2\, R_{\mu\nu}^{(1)} R^{\mu\nu}_{(1)} + \alpha_3\, R_{\mu\nu\lambda\sigma}^{(1)} R^{\mu\nu\lambda\sigma}_{(1)} \right]   \\ 
&=& \int \diff^Dx\,\left[\left( \alpha_1+\frac{\alpha_2}{2}+\alpha_3\right) \partial_\mu \partial_\nu h^{\mu\nu}\partial_\lambda \partial_\sigma h^{\lambda\sigma}+\left( \alpha_1+\frac{\alpha_2}{4}\right) \left( \partial^2 h\partial^2 h-2\partial_\mu \partial_\nu h^{\mu\nu} \partial^2 h \right)\right. \nonumber \\
 &{}& \left.\qquad\qquad+\left(\frac{\alpha_2}{4}+\alpha_3\right) \left( \partial^2 h^{\mu\nu}\partial^2 h_{\mu\nu}-2\partial_\mu \partial_\lambda h^{\mu\sigma}\partial^\nu \partial^\lambda h_{\nu\sigma}\right)\right] \, .\nonumber
\eea
Note that if we set  $\alpha_1=\alpha_3$ and $\alpha_2=-4\alpha_3$ in (\ref{quadi}), the quadratic piece becomes proportional to the Gauss-Bonnet density. For that theory, the third term appearing in the general non-linear equations of motion (\ref{eomG}) is absent, since $\nabla^\alpha P_{\alpha\mu\beta\nu}=0$ for Lovelock theories \cite{Lovelock:1971yv,Padmanabhan:2013xyr}. Hence, following our previous reasoning, the Gauss-Bonnet density should make no contribution whatsoever to the linearized equations on a Minkowki background. This is precisely what happens, as is obvious from $S_{\rm \ssc HD}$, which identically vanishes for that set of couplings.

Rewriting the couplings $\alpha_1$ and $\alpha_2$ in terms of $m_s$ and $m_g$ as
\be
\alpha_1=\frac{(D-2)m_g^2+ D m_s^2 }{4(D-1) m_s^2\, m_g^2}+\alpha_3\, ,\qquad \alpha_2= - \frac{1}{m_g^2} -4\alpha_3\, ,
\ee 
 the original quadratic theory can be written in the following form
\begin{equation}
   \mathcal{L}=\frac{1}{16\pi G} \left[R + \frac{(D-2)(m_g^2-m_s^2)}{4(D-1)m_s^2 m_g^2} R^2 -\frac{(D-2)}{4(D-3)m_g^2} \left[ C_{\mu\nu\rho\sigma}C^{\mu\nu\rho\sigma}- \mathcal{X}_4 \right] + \alpha_3 \mathcal{X}_4  \right]\, , 
\end{equation}
where $C_{\mu\nu\rho\sigma}$ is the Weyl tensor and $\mathcal{X}_4$ is the Gauss-Bonnet density. The equation of motion reads
\begin{align}\label{lineq2}
16\pi G \mathcal{E}_{\mu\nu}^{{(1)}}\equiv  \left[1-\frac{\partial^2}{m_g^2} \right] \left[R_{\mu\nu}^{(1)} -\frac{1}{2}\eta_{\mu\nu}R^{(1)}\right]+ \left[ \frac{(D-2)(m_g^2-m_s^2)}{2(D-1)m_s^2 m_g^2} \right] \left[\eta_{\mu\nu} \partial^2-\partial_{\mu} \partial_{\nu}\right]R^{(1)} \, .
\end{align}
As explained in the main text, the two  coefficients $m_g^2$ and $m_s^2$ are the squared-masses of two extra modes which appear in the linearized spectrum of (\ref{eomG}), in addition to the usual transverse and traceless graviton.
To see this explicitly, it is convenient to rewrite the above expressions in the de Donder gauge,
\begin{equation}
\partial_{\mu}h^{\mu\nu}=\frac{1}{2}\partial^{\nu}h\, .
\end{equation}
In this gauge, the linearized Ricci tensor and Ricci scalar become
\begin{align}
R_{\mu\nu}^{(1)}=-\frac{1}{2}\partial^2 h_{\mu\nu} \, , \quad
R^{(1)}=-\frac{1}{2} \partial^2 h \, ,
\end{align}
and the linearized equations can be written as
\begin{align}\label{lineq4}
\mathcal{E}_{\mu\nu}^{(1)}\equiv -\frac{1}{32\pi G} \partial^2 \hat h_{\mu\nu}=0\, , 
\end{align}
where
\begin{equation}
\hat h_{\mu\nu}\equiv h_{\mu\nu}-\frac{1}{2}\eta_{\mu\nu} h -\frac{1}{m_g^2}\left[\partial^2 h_{\mu\nu}-\frac{1}{2}\partial_{\mu}\partial_{\nu}h \right]+\left[\frac{m_g^2(D-2)+m_s^2}{2(D-1)m_g^2m_s^2} \right] [\eta_{\mu\nu}\partial^2-\partial_{\mu}\partial_{\nu}]h\, .
\end{equation}
This modified perturbation, which corresponds to the usual massless graviton is transverse but not traceless
\begin{equation}
\partial_{\mu}\hat h^{\mu\nu}=0\, , \quad \hat h \neq 0\, .
\end{equation}
Decomposing $h_{\mu\nu}$ as \cite{Bueno:2016ypa}
\begin{equation}
    h_{\mu\nu}= t_{\mu\nu}   +\hat h_{\mu\nu}-\frac{\eta_{\mu\nu}\hat h}{(D-2)} + \frac{(m_g^{-2}-m_s^2)}{(D-1)} \partial_{\langle\mu}\partial_{\nu\rangle}\hat h +\frac{2}{D(D-2)}\eta_{\mu\nu}\phi+\frac{1}{(D-1)m_s^2}  \partial_{\langle \mu}\partial_{\nu \rangle}\phi \, ,
\end{equation}
where $t_{\mu\nu}$ is a traceless spin-2 mode, $\phi$ is a scalar and $\langle \mu \nu \rangle$ denotes the traceless part, it is possible to show that these satisfy
\begin{equation}
    (\partial^2-m_s^2)\phi=0\, , \quad    (\partial^2-m_g^2)t_{\mu\nu}=0\, .
\end{equation}
Hence, they describe massive modes of spin $0$ and $2$ respectively, as anticipated. Had we included some matter stress tensor to the Lagrangian, $t_{\mu\nu}$ would couple to it with the wrong sign, which reflects its ghost-like nature \cite{Alvarez-Gaume:2015rwa,Sisman:2011gz,Bueno:2016ypa}.

\section{Generalized Symmetries of Higher-curvature Gravitons}\label{GSHG}
The generalized symmetries of linearized higher-curvature gravity theories can be characterized using the traceless tensor $\mathcal{W}$ defined as 
\be
\mathcal{W}_{\mu\nu\alpha\beta} \equiv  \mathcal{R}_{\mu\nu\alpha\beta} + \frac{2}{(D-2)} \,\Big[ \eta_{\nu[\alpha} \mathcal{R}_{\beta]\mu}-\eta_{\mu[\alpha} \mathcal{R}_{\beta]\nu}\Big] +  \frac{2}{(D-2)(D-1)}  \eta_{\mu[\alpha}\eta_{\beta]\nu} \, \mathcal{R}\, , \label{Jw2}
\ee
where we have further defined
\be
\mathcal{R}_{\mu\nu\alpha\beta}\equiv \left[1-\frac{\partial^2}{m_g^2} \right]R^{(1)}_{\mu\nu\alpha\beta}+2\Delta_{\mu[\beta}R^{(1)}_{\alpha]\nu}+2\Delta_{\nu[\alpha}R^{(1)}_{\beta]\mu}\, . \label{JJ2}
\ee
The divergence of the biform $\mathcal{W}$ can be computed in any linearized higher-curvature gravity to be 
\be 
\partial^\mu \mathcal{W}_{\mu\nu\alpha\beta}=  \frac{1}{(D-2)} \left[\,\partial_\beta\Big(\mathcal{R}_{\nu\alpha} - \frac{\eta_{\nu\alpha}}{(D-1)}\, \mathcal{R}\Big)-\partial_\alpha \Big(\mathcal{R}_{\nu\beta} - \frac{\eta_{\nu\beta}}{(D-1)}\, \mathcal{R}\Big) \right]\, , \label{divW}
\ee
where the terms inside the parenthesis can be computed using the definition (\ref{JJ2}), the second Bianchi identity, and the equation of motion. They read
\bea \mathcal{R}_{\mu\nu} - \frac{\eta_{\mu\nu}}{(D-1)}\, \mathcal{R} = \frac{(D-2)(m_g^2-m_s^2)}{2(D-1)m_s^2 m_g^2}\left[ (D-3)\,\partial^2 \, \Big( R_{\mu\nu}-\frac{\eta_{\mu\nu}}{2(D-1)} R\Big) -\frac{(D-4)}{2}\partial_\mu\partial_\nu R\right] \, .\label{divW2}\eea
Equation (\ref{divW}) clearly vanishes in two different cases: when $m_g^2=m_s^2$ and when $m_g\to \infty$. Let us start analyzing the latter case. When $m_g\to \infty$ we recover a theory that only contains the usual massless graviton and a massive scalar field. In such case, the equation of motion can be written as 
\be 
R_{\mu\nu}= \frac{1}{2(D-1)} \left[\frac{(D-2)}{m_s^2}\partial_\mu\partial_\nu +\eta_{\mu\nu} \right]  \,R\;,
\ee
which can be used to explicitly check from  (\ref{divW}) and (\ref{divW2}) that $\mathcal{W}$ defines a conserved biform current. However,  in this case, we can further simplify (\ref{Jw2}) to obtain a more amenable expression
\be 
\mathcal{W}_{\mu\nu\alpha\beta} = {R}_{\mu\nu\alpha\beta}+ \frac{1}{(D-1)m_s^2} \Big[ \eta_{\nu[\alpha}\partial_{\beta]}\partial_\mu+\eta_{\mu[\beta}\partial_{\alpha ]}\partial_\nu \Big]\,R\;. \label{ws}
\ee
From (\ref{ws}), one can explicitly  check that  $\mathcal{W}$ obeys the same properties as the Fierz-Pauli Riemann tensor namely
\begin{alignat}{3} \label{SS}
\partial^\mu \,\mathcal{W}_{\mu\nu\alpha\beta}=0  \,,\quad \eta^{\mu\alpha}\, \mathcal{W}_{\mu\nu\alpha\beta}=0 \,,\quad  \varepsilon^{\mu_1...\mu_{D-3}\alpha\beta\gamma}\, \mathcal{W}_{\alpha\beta\gamma\nu}=0
 \,,\quad  \varepsilon^{\mu_1...\mu_{D-3}\alpha\beta\gamma}\,\partial_\alpha\, \mathcal{W}_{\beta\gamma\mu\nu}=0\,.
\end{alignat}
This implies that $\mathcal{W}$  produces non-trivial charges of the type-B and type-D classes. Namely, the ones generated by the conserved currents
\bea
{B}_{\mu\nu}[\mathcal{W}] &=& \mathcal{W}_{\mu\nu\alpha\beta}\,(x^\alpha {b}^\beta-x^\beta {b}^\alpha)\,,\label{BW2}\\
{D}_{\mu\nu} [\mathcal{W}]&=& \mathcal{W}_{\mu\nu\alpha\beta}\,(x^\alpha {d}^{\beta\gamma}x^\gamma-x^\beta {d}^{\alpha\gamma}x^\gamma+\frac{1}{2} {d}^{\alpha\beta} x^2)\,.\label{DW2}
\eea
The remaining charges of the  type-A and type-C classes can be seen to be produced by exact  forms. This is achived by considering same tensor structures as in the Einstein gravity case. To be specific, we obtain that
\bea
{A}_{\mu\nu} [\mathcal{W}] &=& \mathcal{W}_{\mu\nu\alpha\beta} \, {a}^{\alpha\beta} =\partial^\rho \mathcal{A}_{\mu\nu\rho}[\mathcal{W}]\,,\label{AW2}\\
{C}_{\mu\nu} [\mathcal{W}] &=& \mathcal{W}_{\mu\nu\alpha\beta}\,{c}^{\alpha\beta\gamma}x_\gamma=\partial^\rho \mathcal{C}_{\mu\nu\rho}[\mathcal{W}]\,.\label{CW2} 
\eea
where we have used the notation
\bea \label{AtriW}
\mathcal{A}_{\mu\nu\rho} [\mathcal{W}] &\equiv &- \frac{\mathcal{W}^*_{\mu\nu\rho\alpha_1 ...\alpha_{D-3}}}{(D-4)!}\,\tilde{a}^{\alpha_1 ...\alpha_{D-3}\sigma}\,x_\sigma\,,\\
\mathcal{C}_{\mu\nu\rho} [\mathcal{W}] &\equiv & \frac{\mathcal{W}^*_{\mu\nu\rho\alpha_1 ...\alpha_{D-3}}}{(D-5)!(D-2)}\,\left(\,\frac{1}{2}\,\tilde{c}^{\alpha_1 ...\alpha_{D-3}}\,x^2+\, \frac{\eta^{\alpha_1 ...\alpha_{D-3}}_{\beta_1 ...\beta_{D-3}}}{(D-4)!}\, c^{\beta_1 ...\beta_{D-4}\sigma} \,x^{\beta_{D-3} }\, x_\sigma\right)\;. \label{exactW}
\eea
Following this line, an issue we have avoided so far is that $\mathcal{R}$, defined from (\ref{JJ2}), is by itself a conserved biform current in any higher-order gravity theory, namely $\partial^{\mu}\mathcal{R}_{\mu\nu\alpha\beta}=0$.  However,  it has a non-vanishing trace producing only charges of the type-A and type-C classes. We can check that these are generated by exact currents. For instance, we obtain that in the limit $m_g\to \infty$, 
\be
 {A}_{\mu\nu} [\mathcal{R}]- {A}_{\mu\nu} [\mathcal{W}] =\partial^\rho \left(\frac{\mathcal{A}_{\mu\nu\rho}[\mathcal{W}] -\mathcal{A}_{\mu\nu\rho}[R] }{(D-3)(D-1) }\right) \;,
\ee
where we have used that (\ref{JJ2}) reduces to
\bea 
\mathcal{R}_{\mu\nu\alpha\beta} = {R}_{\mu\nu\alpha\beta} &+& \frac{(D-2)}{(D-1)^2 m_s^2} \Big[ \eta_{\nu[\alpha}\partial_{\beta]}\partial_\mu +\eta_{\mu[\beta}\partial_{\alpha ]}\partial_\nu  \Big]R - \frac{1}{(D-1)^2} \eta_{\mu[\alpha}\eta_{\beta]\nu} \, R\;.
\eea
It rests to check what happens in the dual picture, where the dual tensor  $\mathcal{W}^*_{\mu_1 ...\mu_{D-2}\alpha\beta}$ is also conserved and generates charges of type-$\tilde{A}$ and  type-$\tilde{B}$ in the usual manner
\bea 
\tilde{A}_{\mu_1 \mu_2...\mu_{D-2}} [\mathcal{W}] &=& \mathcal{W}^*_{\mu_1 \mu_2...\mu_{D-2}\alpha\beta} \, \tilde{a}^{\alpha\beta}\,,\label{ATd}\\
\tilde{B}_{\mu_1 \mu_2...\mu_{D-2}} [\mathcal{W}]&=& \mathcal{W}^*_{\mu_1 \mu_2...\mu_{D-2}\alpha\beta}\,(x^\alpha \tilde{b}^\beta-x^\beta \tilde{b}^\alpha)\,.\label{BTd}
\eea
Notice that this dual tensor can be written as 
\be 
\mathcal{W}^*_{\mu_1 ...\mu_{D-2}\alpha\beta} = R^*_{\mu_1 ...\mu_{D-2}\alpha\beta}+\frac{1}{(D-1)} \varepsilon_{\mu_1 ...\mu_{D-2}\rho\sigma} \eta^{\rho}_{[\beta} \partial_{\alpha ]}\partial^\sigma \, R\;,
\ee
which implies that $\mathcal{W}^*_{\mu_1 ...\mu_{D-2}\alpha\beta}$ generates the same charges as $R^*_{\mu_1 ...\mu_{D-2}\alpha\beta}$, the reason being the remaining pieces can be seen to be exact too. This follows by defining
\bea 
\tilde{\mathcal{A}}_{\mu_1 ...\mu_{D-1}}&\equiv &\frac{1}{(D-1)} \varepsilon_{\mu_1 ...\mu_{D-1}\rho}\, a^{\alpha\beta}\,\eta^{\rho}_{[\alpha} \partial_{\beta ]}R \;,\\
\tilde{\mathcal{B}}_{\mu_1 ...\mu_{D-1}}&\equiv &\frac{1}{(D-1)} \Big[\varepsilon_{\mu_1 ...\mu_{D-1}\rho}\, \big(b^{\alpha}x^{\beta}-b^{\beta}x^{\alpha}\big)\,\eta^{\rho}_{[\alpha} \partial_{\beta ]} R - \, \varepsilon_{\mu_1 ...\mu_{D-1}\rho}b^\rho \, R\Big]\;,
\eea
whose divergences yield
\bea 
\partial^\sigma\tilde{\mathcal{A}}_{\mu_1 ...\mu_{D-2}\sigma}&=& \frac{1}{(D-1)} \varepsilon_{\mu_1 ...\mu_{D-2}\rho\sigma}\, a^{\alpha\beta}\,\eta^{\rho}_{[\beta} \partial_{\alpha ]}\partial^\sigma R= \tilde{A}_{\mu_1 \mu_2...\mu_{D-2}} [\mathcal{W}] - \tilde{A}_{\mu_1 \mu_2...\mu_{D-2}} [R^*];,\\
\partial^\sigma\tilde{\mathcal{B}}_{\mu_1 ...\mu_{D-2}\sigma}&=& \frac{1}{(D-1)} \varepsilon_{\mu_1 ...\mu_{D-2}\rho\sigma}\, \Big(b^{\alpha}x^{\beta}-b^{\beta}x^{\alpha}\Big)\,\eta^{\rho}_{[\beta} \partial_{\alpha ]} \partial^\sigma R = \tilde{B}_{\mu_1 \mu_2...\mu_{D-2}} [\mathcal{W}] - \tilde{B}_{\mu_1 \mu_2...\mu_{D-2}} [R^*]\;.
\eea
Finally, we are left to analyze the case $m^2_s=m^2_g=m^2$ whose logic follows a similar pattern. In such case, the equation of motion can be simplified to be 
\be 
\left[1-\frac{\partial^2}{m^2}\right]R_{\mu\nu} = 0\;.
\ee
In addition, equations (\ref{Jw2}) and (\ref{JJ2}) become
\be
\mathcal{W}_{\mu\nu\alpha\beta} = \mathcal{R}_{\mu\nu\alpha\beta} = \left[1-\frac{\partial^2}{m^2}\right] R_{\mu\nu\alpha\beta}\;,
\ee
which obeys all the properties presented in (\ref{SS}), therefore generating non-trivial charges from the currents (\ref{BW2}), (\ref{DW2}), (\ref{ATd}) and (\ref{BTd}). The corresponding calculations and proofs  showing the remaining conserved currents are exact are analogous to the ones already presented.   
\section{Generalized symmetries define the generalized graviton}
\label{gravfromsym}
In \cite{hofman} an interesting perspective was taken in relation to gravity. Following the Landau paradigm \cite{alma991009084709703276} ---and in particular its generalization to include generalized symmetries \cite{Gaiotto:2014kfa}--- one characterizes a phase of matter (a theory) in terms of its symmetries and its pattern of symmetry breaking. Gravity might follow this paradigm, and indeed this was shown for the linearized Einstein graviton in \cite{hofman}, where the Fierz-Pauli action was rederived from the symmetry pattern. Here we show that this is the case for generalized gravitons as well. Instead of focusing on re-deriving the appropriate action, we focus on re-deriving the appropriate equations of motion, namely the full set of linearized  equations for generalized gravitons, from the symmetry pattern of the theory.

It is a straightforward exercise to see that, assuming we look for a theory with the generalized symmetries generated by the usual families
\be 
Q= \int_{\Sigma_{2}}\,\left(\,\tilde{A}\,+\,\tilde{B}\,\right)\,, \label{DCho}
\ee
and their dual $B,D$ families, then the closedness of these families implies the full set of Einstein equations. For example, for the Einstein graviton  we aim to recover (\ref{EE}-\ref{sbi}). This can be seen to follow from
\be
\diff \,\star B=0\,,\quad\,\,\,\,\,\diff \,\star D=0\,,\quad\,\,\,\,\,\diff \,\star \tilde{A}=0\,,\quad\,\,\,\,\,\diff \,\star \tilde{B}=0\;,
\ee
which enforce the Bianchi identities of the Riemann tensor to hold  as well as the Einstein equations. For generalized gravitons we must consider the same closedness conditions but remembering that it is the $\mathcal{W}$ tensor the one generating the charges. Then, the conservation of $\tilde{A}$ implies the validity of the second Bianchi identity (\ref{sbi}). The closedness of $B$ sets the equations of motion (\ref{lineq2}) to zero. Finally, the behavior of $D$ and $\tilde{B}$ enforce the first Bianchi identity (\ref{FB}). Therefore, we conclude that the full set of  equations of motion for general theories of linearized gravity follows from  their generalized symmetries pattern.

\bibliography{biblio}

\begin{thebibliography}{66}%
\makeatletter
\providecommand \@ifxundefined [1]{%
 \@ifx{#1\undefined}
}%
\providecommand \@ifnum [1]{%
 \ifnum #1\expandafter \@firstoftwo
 \else \expandafter \@secondoftwo
 \fi
}%
\providecommand \@ifx [1]{%
 \ifx #1\expandafter \@firstoftwo
 \else \expandafter \@secondoftwo
 \fi
}%
\providecommand \natexlab [1]{#1}%
\providecommand \enquote  [1]{``#1''}%
\providecommand \bibnamefont  [1]{#1}%
\providecommand \bibfnamefont [1]{#1}%
\providecommand \citenamefont [1]{#1}%
\providecommand \href@noop [0]{\@secondoftwo}%
\providecommand \href [0]{\begingroup \@sanitize@url \@href}%
\providecommand \@href[1]{\@@startlink{#1}\@@href}%
\providecommand \@@href[1]{\endgroup#1\@@endlink}%
\providecommand \@sanitize@url [0]{\catcode `\\12\catcode `\$12\catcode
  `\&12\catcode `\#12\catcode `\^12\catcode `\_12\catcode `\%12\relax}%
\providecommand \@@startlink[1]{}%
\providecommand \@@endlink[0]{}%
\providecommand \url  [0]{\begingroup\@sanitize@url \@url }%
\providecommand \@url [1]{\endgroup\@href {#1}{\urlprefix }}%
\providecommand \urlprefix  [0]{URL }%
\providecommand \Eprint [0]{\href }%
\providecommand \doibase [0]{http://dx.doi.org/}%
\providecommand \selectlanguage [0]{\@gobble}%
\providecommand \bibinfo  [0]{\@secondoftwo}%
\providecommand \bibfield  [0]{\@secondoftwo}%
\providecommand \translation [1]{[#1]}%
\providecommand \BibitemOpen [0]{}%
\providecommand \bibitemStop [0]{}%
\providecommand \bibitemNoStop [0]{.\EOS\space}%
\providecommand \EOS [0]{\spacefactor3000\relax}%
\providecommand \BibitemShut  [1]{\csname bibitem#1\endcsname}%
\let\auto@bib@innerbib\@empty
\bibitem [{\citenamefont {'t~Hooft}(1978)}]{tHooft:1977nqb}%
  \BibitemOpen
  \bibfield  {author} {\bibinfo {author} {\bibfnamefont {G.}~\bibnamefont
  {'t~Hooft}},\ }\href {\doibase 10.1016/0550-3213(78)90153-0} {\bibfield
  {journal} {\bibinfo  {journal} {Nucl. Phys. B}\ }\textbf {\bibinfo {volume}
  {138}},\ \bibinfo {pages} {1} (\bibinfo {year} {1978})}\BibitemShut {NoStop}%
\bibitem [{\citenamefont {Gaiotto}\ \emph {et~al.}(2015)\citenamefont
  {Gaiotto}, \citenamefont {Kapustin}, \citenamefont {Seiberg},\ and\
  \citenamefont {Willett}}]{Gaiotto:2014kfa}%
  \BibitemOpen
  \bibfield  {author} {\bibinfo {author} {\bibfnamefont {D.}~\bibnamefont
  {Gaiotto}}, \bibinfo {author} {\bibfnamefont {A.}~\bibnamefont {Kapustin}},
  \bibinfo {author} {\bibfnamefont {N.}~\bibnamefont {Seiberg}}, \ and\
  \bibinfo {author} {\bibfnamefont {B.}~\bibnamefont {Willett}},\ }\href
  {\doibase 10.1007/JHEP02(2015)172} {\bibfield  {journal} {\bibinfo  {journal}
  {JHEP}\ }\textbf {\bibinfo {volume} {02}},\ \bibinfo {pages} {172} (\bibinfo
  {year} {2015})},\ \Eprint {http://arxiv.org/abs/1412.5148} {arXiv:1412.5148
  [hep-th]} \BibitemShut {NoStop}%
\bibitem [{\citenamefont {Cordova}\ \emph {et~al.}(2022)\citenamefont
  {Cordova}, \citenamefont {Dumitrescu}, \citenamefont {Intriligator},\ and\
  \citenamefont {Shao}}]{Cordova:2022ruw}%
  \BibitemOpen
  \bibfield  {author} {\bibinfo {author} {\bibfnamefont {C.}~\bibnamefont
  {Cordova}}, \bibinfo {author} {\bibfnamefont {T.~T.}\ \bibnamefont
  {Dumitrescu}}, \bibinfo {author} {\bibfnamefont {K.}~\bibnamefont
  {Intriligator}}, \ and\ \bibinfo {author} {\bibfnamefont {S.-H.}\
  \bibnamefont {Shao}},\ }in\ \href@noop {} {\emph {\bibinfo {booktitle} {{2022
  Snowmass Summer Study}}}}\ (\bibinfo {year} {2022})\ \Eprint
  {http://arxiv.org/abs/2205.09545} {arXiv:2205.09545 [hep-th]} \BibitemShut
  {NoStop}%
\bibitem [{\citenamefont {McGreevy}(2022)}]{McGreevy:2022oyu}%
  \BibitemOpen
  \bibfield  {author} {\bibinfo {author} {\bibfnamefont {J.}~\bibnamefont
  {McGreevy}},\ }\href@noop {} {\  (\bibinfo {year} {2022})},\ \Eprint
  {http://arxiv.org/abs/2204.03045} {arXiv:2204.03045 [cond-mat.str-el]}
  \BibitemShut {NoStop}%
\bibitem [{\citenamefont {Casini}\ and\ \citenamefont
  {Magan}(2021)}]{Casini:2021zgr}%
  \BibitemOpen
  \bibfield  {author} {\bibinfo {author} {\bibfnamefont {H.}~\bibnamefont
  {Casini}}\ and\ \bibinfo {author} {\bibfnamefont {J.~M.}\ \bibnamefont
  {Magan}},\ }\href {\doibase 10.1142/S0217732321300251} {\bibfield  {journal}
  {\bibinfo  {journal} {Mod. Phys. Lett. A}\ }\textbf {\bibinfo {volume}
  {36}},\ \bibinfo {pages} {2130025} (\bibinfo {year} {2021})},\ \Eprint
  {http://arxiv.org/abs/2110.11358} {arXiv:2110.11358 [hep-th]} \BibitemShut
  {NoStop}%
\bibitem [{\citenamefont {Fierz}\ and\ \citenamefont
  {Pauli}(1939)}]{Fierz:1939ix}%
  \BibitemOpen
  \bibfield  {author} {\bibinfo {author} {\bibfnamefont {M.}~\bibnamefont
  {Fierz}}\ and\ \bibinfo {author} {\bibfnamefont {W.}~\bibnamefont {Pauli}},\
  }\href {\doibase 10.1098/rspa.1939.0140} {\bibfield  {journal} {\bibinfo
  {journal} {Proc. Roy. Soc. Lond. A}\ }\textbf {\bibinfo {volume} {173}},\
  \bibinfo {pages} {211} (\bibinfo {year} {1939})}\BibitemShut {NoStop}%
\bibitem [{\citenamefont {Benedetti}\ \emph
  {et~al.}(2022{\natexlab{a}})\citenamefont {Benedetti}, \citenamefont
  {Casini},\ and\ \citenamefont {Magan}}]{casini2021generalized}%
  \BibitemOpen
  \bibfield  {author} {\bibinfo {author} {\bibfnamefont {V.}~\bibnamefont
  {Benedetti}}, \bibinfo {author} {\bibfnamefont {H.}~\bibnamefont {Casini}}, \
  and\ \bibinfo {author} {\bibfnamefont {J.~M.}\ \bibnamefont {Magan}},\ }\href
  {\doibase 10.1007/JHEP05(2022)045} {\bibfield  {journal} {\bibinfo  {journal}
  {JHEP}\ }\textbf {\bibinfo {volume} {05}},\ \bibinfo {pages} {045} (\bibinfo
  {year} {2022}{\natexlab{a}})},\ \Eprint {http://arxiv.org/abs/2111.12089}
  {arXiv:2111.12089 [hep-th]} \BibitemShut {NoStop}%
\bibitem [{\citenamefont {Benedetti}\ \emph
  {et~al.}(2022{\natexlab{b}})\citenamefont {Benedetti}, \citenamefont
  {Casini},\ and\ \citenamefont {Magan}}]{Benedetti:2022zbb}%
  \BibitemOpen
  \bibfield  {author} {\bibinfo {author} {\bibfnamefont {V.}~\bibnamefont
  {Benedetti}}, \bibinfo {author} {\bibfnamefont {H.}~\bibnamefont {Casini}}, \
  and\ \bibinfo {author} {\bibfnamefont {J.~M.}\ \bibnamefont {Magan}},\ }\href
  {\doibase 10.1007/JHEP08(2022)304} {\bibfield  {journal} {\bibinfo  {journal}
  {JHEP}\ }\textbf {\bibinfo {volume} {08}},\ \bibinfo {pages} {304} (\bibinfo
  {year} {2022}{\natexlab{b}})},\ \Eprint {http://arxiv.org/abs/2205.03412}
  {arXiv:2205.03412 [hep-th]} \BibitemShut {NoStop}%
\bibitem [{\citenamefont {Hinterbichler}\ \emph {et~al.}(2023)\citenamefont
  {Hinterbichler}, \citenamefont {Hofman}, \citenamefont {Joyce},\ and\
  \citenamefont {Mathys}}]{hofman}%
  \BibitemOpen
  \bibfield  {author} {\bibinfo {author} {\bibfnamefont {K.}~\bibnamefont
  {Hinterbichler}}, \bibinfo {author} {\bibfnamefont {D.~M.}\ \bibnamefont
  {Hofman}}, \bibinfo {author} {\bibfnamefont {A.}~\bibnamefont {Joyce}}, \
  and\ \bibinfo {author} {\bibfnamefont {G.}~\bibnamefont {Mathys}},\ }\href
  {\doibase 10.1007/JHEP02(2023)151} {\bibfield  {journal} {\bibinfo  {journal}
  {JHEP}\ }\textbf {\bibinfo {volume} {02}},\ \bibinfo {pages} {151} (\bibinfo
  {year} {2023})},\ \Eprint {http://arxiv.org/abs/2205.12272} {arXiv:2205.12272
  [hep-th]} \BibitemShut {NoStop}%
\bibitem [{\citenamefont {Pretko}(2017{\natexlab{a}})}]{Pretko:2017}%
  \BibitemOpen
  \bibfield  {author} {\bibinfo {author} {\bibfnamefont {M.}~\bibnamefont
  {Pretko}},\ }\href {\doibase 10.1103/PhysRevD.96.024051} {\bibfield
  {journal} {\bibinfo  {journal} {Phys. Rev. D}\ }\textbf {\bibinfo {volume}
  {96}},\ \bibinfo {pages} {024051} (\bibinfo {year} {2017}{\natexlab{a}})},\
  \Eprint {http://arxiv.org/abs/1702.07613} {arXiv:1702.07613
  [cond-mat.str-el]} \BibitemShut {NoStop}%
\bibitem [{\citenamefont {Lee}\ and\ \citenamefont {Wald}(1990)}]{Lee:1990nz}%
  \BibitemOpen
  \bibfield  {author} {\bibinfo {author} {\bibfnamefont {J.}~\bibnamefont
  {Lee}}\ and\ \bibinfo {author} {\bibfnamefont {R.~M.}\ \bibnamefont {Wald}},\
  }\href {\doibase 10.1063/1.528801} {\bibfield  {journal} {\bibinfo  {journal}
  {J. Math. Phys.}\ }\textbf {\bibinfo {volume} {31}},\ \bibinfo {pages} {725}
  (\bibinfo {year} {1990})}\BibitemShut {NoStop}%
\bibitem [{\citenamefont {Wald}(1993)}]{Wald:1993nt}%
  \BibitemOpen
  \bibfield  {author} {\bibinfo {author} {\bibfnamefont {R.~M.}\ \bibnamefont
  {Wald}},\ }\href {\doibase 10.1103/PhysRevD.48.R3427} {\bibfield  {journal}
  {\bibinfo  {journal} {Phys. Rev. D}\ }\textbf {\bibinfo {volume} {48}},\
  \bibinfo {pages} {R3427} (\bibinfo {year} {1993})},\ \Eprint
  {http://arxiv.org/abs/gr-qc/9307038} {arXiv:gr-qc/9307038} \BibitemShut
  {NoStop}%
\bibitem [{\citenamefont {Iyer}\ and\ \citenamefont {Wald}(1994)}]{Iyer_1994}%
  \BibitemOpen
  \bibfield  {author} {\bibinfo {author} {\bibfnamefont {V.}~\bibnamefont
  {Iyer}}\ and\ \bibinfo {author} {\bibfnamefont {R.~M.}\ \bibnamefont
  {Wald}},\ }\href {\doibase 10.1103/physrevd.50.846} {\bibfield  {journal}
  {\bibinfo  {journal} {Physical Review D}\ }\textbf {\bibinfo {volume} {50}},\
  \bibinfo {pages} {846} (\bibinfo {year} {1994})}\BibitemShut {NoStop}%
\bibitem [{\citenamefont {Casini}\ \emph {et~al.}(2021)\citenamefont {Casini},
  \citenamefont {Huerta}, \citenamefont {Magan},\ and\ \citenamefont
  {Pontello}}]{Casini:2020rgj}%
  \BibitemOpen
  \bibfield  {author} {\bibinfo {author} {\bibfnamefont {H.}~\bibnamefont
  {Casini}}, \bibinfo {author} {\bibfnamefont {M.}~\bibnamefont {Huerta}},
  \bibinfo {author} {\bibfnamefont {J.~M.}\ \bibnamefont {Magan}}, \ and\
  \bibinfo {author} {\bibfnamefont {D.}~\bibnamefont {Pontello}},\ }\href
  {\doibase 10.1007/JHEP04(2021)277} {\bibfield  {journal} {\bibinfo  {journal}
  {JHEP}\ }\textbf {\bibinfo {volume} {04}},\ \bibinfo {pages} {277} (\bibinfo
  {year} {2021})},\ \Eprint {http://arxiv.org/abs/2008.11748} {arXiv:2008.11748
  [hep-th]} \BibitemShut {NoStop}%
\bibitem [{\citenamefont {Magan}\ and\ \citenamefont
  {Pontello}(2021)}]{Magan:2020ake}%
  \BibitemOpen
  \bibfield  {author} {\bibinfo {author} {\bibfnamefont {J.~M.}\ \bibnamefont
  {Magan}}\ and\ \bibinfo {author} {\bibfnamefont {D.}~\bibnamefont
  {Pontello}},\ }\href {\doibase 10.1103/PhysRevA.103.012211} {\bibfield
  {journal} {\bibinfo  {journal} {Phys. Rev. A}\ }\textbf {\bibinfo {volume}
  {103}},\ \bibinfo {pages} {012211} (\bibinfo {year} {2021})},\ \Eprint
  {http://arxiv.org/abs/2005.01760} {arXiv:2005.01760 [hep-th]} \BibitemShut
  {NoStop}%
\bibitem [{\citenamefont {Casini}\ \emph {et~al.}(2020)\citenamefont {Casini},
  \citenamefont {Huerta}, \citenamefont {Magán},\ and\ \citenamefont
  {Pontello}}]{Casini:2019kex}%
  \BibitemOpen
  \bibfield  {author} {\bibinfo {author} {\bibfnamefont {H.}~\bibnamefont
  {Casini}}, \bibinfo {author} {\bibfnamefont {M.}~\bibnamefont {Huerta}},
  \bibinfo {author} {\bibfnamefont {J.~M.}\ \bibnamefont {Magán}}, \ and\
  \bibinfo {author} {\bibfnamefont {D.}~\bibnamefont {Pontello}},\ }\href
  {\doibase 10.1007/JHEP02(2020)014} {\bibfield  {journal} {\bibinfo  {journal}
  {JHEP}\ }\textbf {\bibinfo {volume} {02}},\ \bibinfo {pages} {014} (\bibinfo
  {year} {2020})},\ \Eprint {http://arxiv.org/abs/1905.10487} {arXiv:1905.10487
  [hep-th]} \BibitemShut {NoStop}%
\bibitem [{\citenamefont {Magan}(2021)}]{Magan:2021myk}%
  \BibitemOpen
  \bibfield  {author} {\bibinfo {author} {\bibfnamefont {J.~M.}\ \bibnamefont
  {Magan}},\ }\href {\doibase 10.1007/JHEP12(2021)100} {\bibfield  {journal}
  {\bibinfo  {journal} {JHEP}\ }\textbf {\bibinfo {volume} {12}},\ \bibinfo
  {pages} {100} (\bibinfo {year} {2021})},\ \Eprint
  {http://arxiv.org/abs/2111.02418} {arXiv:2111.02418 [hep-th]} \BibitemShut
  {NoStop}%
\bibitem [{\citenamefont {Harlow}\ and\ \citenamefont
  {Ooguri}(2021)}]{Harlow:2021trr}%
  \BibitemOpen
  \bibfield  {author} {\bibinfo {author} {\bibfnamefont {D.}~\bibnamefont
  {Harlow}}\ and\ \bibinfo {author} {\bibfnamefont {H.}~\bibnamefont
  {Ooguri}},\ }\href@noop {} {\  (\bibinfo {year} {2021})},\ \Eprint
  {http://arxiv.org/abs/2109.03838} {arXiv:2109.03838 [hep-th]} \BibitemShut
  {NoStop}%
\bibitem [{\citenamefont {Polchinski}(2004)}]{Polchinski:2003bq}%
  \BibitemOpen
  \bibfield  {author} {\bibinfo {author} {\bibfnamefont {J.}~\bibnamefont
  {Polchinski}},\ }\href {\doibase 10.1142/S0217751X0401866X} {\bibfield
  {journal} {\bibinfo  {journal} {Int. J. Mod. Phys. A}\ }\textbf {\bibinfo
  {volume} {19S1}},\ \bibinfo {pages} {145} (\bibinfo {year} {2004})},\ \Eprint
  {http://arxiv.org/abs/hep-th/0304042} {arXiv:hep-th/0304042} \BibitemShut
  {NoStop}%
\bibitem [{\citenamefont {Banks}\ and\ \citenamefont
  {Seiberg}(2011)}]{Banks:2010zn}%
  \BibitemOpen
  \bibfield  {author} {\bibinfo {author} {\bibfnamefont {T.}~\bibnamefont
  {Banks}}\ and\ \bibinfo {author} {\bibfnamefont {N.}~\bibnamefont
  {Seiberg}},\ }\href {\doibase 10.1103/PhysRevD.83.084019} {\bibfield
  {journal} {\bibinfo  {journal} {Phys. Rev. D}\ }\textbf {\bibinfo {volume}
  {83}},\ \bibinfo {pages} {084019} (\bibinfo {year} {2011})},\ \Eprint
  {http://arxiv.org/abs/1011.5120} {arXiv:1011.5120 [hep-th]} \BibitemShut
  {NoStop}%
\bibitem [{\citenamefont {Rudelius}\ and\ \citenamefont
  {Shao}(2020)}]{Rudelius:2020orz}%
  \BibitemOpen
  \bibfield  {author} {\bibinfo {author} {\bibfnamefont {T.}~\bibnamefont
  {Rudelius}}\ and\ \bibinfo {author} {\bibfnamefont {S.-H.}\ \bibnamefont
  {Shao}},\ }\href {\doibase 10.1007/JHEP12(2020)172} {\bibfield  {journal}
  {\bibinfo  {journal} {JHEP}\ }\textbf {\bibinfo {volume} {12}},\ \bibinfo
  {pages} {172} (\bibinfo {year} {2020})},\ \Eprint
  {http://arxiv.org/abs/2006.10052} {arXiv:2006.10052 [hep-th]} \BibitemShut
  {NoStop}%
\bibitem [{\citenamefont {Heidenreich}\ \emph {et~al.}(2020)\citenamefont
  {Heidenreich}, \citenamefont {McNamara}, \citenamefont {Montero},
  \citenamefont {Reece}, \citenamefont {Rudelius},\ and\ \citenamefont
  {Valenzuela}}]{Heidenreich:2020tzg}%
  \BibitemOpen
  \bibfield  {author} {\bibinfo {author} {\bibfnamefont {B.}~\bibnamefont
  {Heidenreich}}, \bibinfo {author} {\bibfnamefont {J.}~\bibnamefont
  {McNamara}}, \bibinfo {author} {\bibfnamefont {M.}~\bibnamefont {Montero}},
  \bibinfo {author} {\bibfnamefont {M.}~\bibnamefont {Reece}}, \bibinfo
  {author} {\bibfnamefont {T.}~\bibnamefont {Rudelius}}, \ and\ \bibinfo
  {author} {\bibfnamefont {I.}~\bibnamefont {Valenzuela}},\ }\href {\doibase
  10.1007/JHEP09(2021)203} {\bibfield  {journal} {\bibinfo  {journal} {JHEP}\
  }\textbf {\bibinfo {volume} {21}},\ \bibinfo {pages} {203} (\bibinfo {year}
  {2020})},\ \Eprint {http://arxiv.org/abs/2104.07036} {arXiv:2104.07036
  [hep-th]} \BibitemShut {NoStop}%
\bibitem [{\citenamefont {Witten}(2023)}]{Witten:2023qsv}%
  \BibitemOpen
  \bibfield  {author} {\bibinfo {author} {\bibfnamefont {E.}~\bibnamefont
  {Witten}},\ }\href@noop {} {\  (\bibinfo {year} {2023})},\ \Eprint
  {http://arxiv.org/abs/2303.02837} {arXiv:2303.02837 [hep-th]} \BibitemShut
  {NoStop}%
\bibitem [{\citenamefont {Pretko}\ \emph {et~al.}(2020)\citenamefont {Pretko},
  \citenamefont {Chen},\ and\ \citenamefont {You}}]{Pretko:2020}%
  \BibitemOpen
  \bibfield  {author} {\bibinfo {author} {\bibfnamefont {M.}~\bibnamefont
  {Pretko}}, \bibinfo {author} {\bibfnamefont {X.}~\bibnamefont {Chen}}, \ and\
  \bibinfo {author} {\bibfnamefont {Y.}~\bibnamefont {You}},\ }\href {\doibase
  10.1142/S0217751X20300033} {\bibfield  {journal} {\bibinfo  {journal} {Int.
  J. Mod. Phys. A}\ }\textbf {\bibinfo {volume} {35}},\ \bibinfo {pages}
  {2030003} (\bibinfo {year} {2020})},\ \Eprint
  {http://arxiv.org/abs/2001.01722} {arXiv:2001.01722 [cond-mat.str-el]}
  \BibitemShut {NoStop}%
\bibitem [{\citenamefont {Rasmussen}\ \emph {et~al.}(2016)\citenamefont
  {Rasmussen}, \citenamefont {You},\ and\ \citenamefont {Xu}}]{Rasmussen2016}%
  \BibitemOpen
  \bibfield  {author} {\bibinfo {author} {\bibfnamefont {A.}~\bibnamefont
  {Rasmussen}}, \bibinfo {author} {\bibfnamefont {Y.-Z.}\ \bibnamefont {You}},
  \ and\ \bibinfo {author} {\bibfnamefont {C.}~\bibnamefont {Xu}},\ }\href@noop
  {} {\  (\bibinfo {year} {2016})},\ \Eprint {http://arxiv.org/abs/1601.08235}
  {arXiv:1601.08235 [cond-mat.str-el]} \BibitemShut {NoStop}%
\bibitem [{\citenamefont {Jones}(1983)}]{Jones1983}%
  \BibitemOpen
  \bibfield  {author} {\bibinfo {author} {\bibfnamefont {V.}~\bibnamefont
  {Jones}},\ }\href {http://eudml.org/doc/143011} {\bibfield  {journal}
  {\bibinfo  {journal} {Inventiones mathematicae}\ }\textbf {\bibinfo {volume}
  {72}},\ \bibinfo {pages} {1} (\bibinfo {year} {1983})}\BibitemShut {NoStop}%
\bibitem [{\citenamefont {Kosaki}(1986)}]{KOSAKI1986123}%
  \BibitemOpen
  \bibfield  {author} {\bibinfo {author} {\bibfnamefont {H.}~\bibnamefont
  {Kosaki}},\ }\href {\doibase https://doi.org/10.1016/0022-1236(86)90085-6}
  {\bibfield  {journal} {\bibinfo  {journal} {Journal of Functional Analysis}\
  }\textbf {\bibinfo {volume} {66}},\ \bibinfo {pages} {123 } (\bibinfo {year}
  {1986})}\BibitemShut {NoStop}%
\bibitem [{\citenamefont {Longo}(1989)}]{longo1989}%
  \BibitemOpen
  \bibfield  {author} {\bibinfo {author} {\bibfnamefont {R.}~\bibnamefont
  {Longo}},\ }\href {https://projecteuclid.org:443/euclid.cmp/1104179850}
  {\bibfield  {journal} {\bibinfo  {journal} {Comm. Math. Phys.}\ }\textbf
  {\bibinfo {volume} {126}},\ \bibinfo {pages} {217} (\bibinfo {year}
  {1989})}\BibitemShut {NoStop}%
\bibitem [{\citenamefont {Teruya}(1992)}]{teruya}%
  \BibitemOpen
  \bibfield  {author} {\bibinfo {author} {\bibfnamefont {T.}~\bibnamefont
  {Teruya}},\ }\href@noop {} {\bibfield  {journal} {\bibinfo  {journal} {Publ.
  Res. Inst. Math. Sci.}\ }\textbf {\bibinfo {volume} {28}},\ \bibinfo {pages}
  {437–453} (\bibinfo {year} {1992})}\BibitemShut {NoStop}%
\bibitem [{\citenamefont {Giorgetti}\ and\ \citenamefont
  {Longo}(2019)}]{giorlongo}%
  \BibitemOpen
  \bibfield  {author} {\bibinfo {author} {\bibfnamefont {L.}~\bibnamefont
  {Giorgetti}}\ and\ \bibinfo {author} {\bibfnamefont {R.}~\bibnamefont
  {Longo}},\ }\href@noop {} {\bibfield  {journal} {\bibinfo  {journal}
  {{Communications in Mathematical Physics}}\ }\textbf {\bibinfo {volume}
  {370}},\ \bibinfo {pages} {719} (\bibinfo {year} {2019})}\BibitemShut
  {NoStop}%
\bibitem [{\citenamefont {Longo}\ \emph {et~al.}(2019)\citenamefont {Longo},
  \citenamefont {Morinelli}, \citenamefont {Preta},\ and\ \citenamefont
  {Rehren}}]{longo2019split}%
  \BibitemOpen
  \bibfield  {author} {\bibinfo {author} {\bibfnamefont {R.}~\bibnamefont
  {Longo}}, \bibinfo {author} {\bibfnamefont {V.}~\bibnamefont {Morinelli}},
  \bibinfo {author} {\bibfnamefont {F.}~\bibnamefont {Preta}}, \ and\ \bibinfo
  {author} {\bibfnamefont {K.-H.}\ \bibnamefont {Rehren}},\ }in\ \href@noop {}
  {\emph {\bibinfo {booktitle} {Annales Henri Poincar{\'e}}}},\ Vol.\ \bibinfo
  {volume} {20 8}\ (\bibinfo {organization} {Springer},\ \bibinfo {year}
  {2019})\ pp.\ \bibinfo {pages} {2555--2584}\BibitemShut {NoStop}%
\bibitem [{\citenamefont {Hull}(2001)}]{Hull:2001iu}%
  \BibitemOpen
  \bibfield  {author} {\bibinfo {author} {\bibfnamefont {C.~M.}\ \bibnamefont
  {Hull}},\ }\href {\doibase 10.1088/1126-6708/2001/09/027} {\bibfield
  {journal} {\bibinfo  {journal} {JHEP}\ }\textbf {\bibinfo {volume} {09}},\
  \bibinfo {pages} {027} (\bibinfo {year} {2001})},\ \Eprint
  {http://arxiv.org/abs/hep-th/0107149} {arXiv:hep-th/0107149} \BibitemShut
  {NoStop}%
\bibitem [{\citenamefont {Henneaux}\ \emph {et~al.}(2020)\citenamefont
  {Henneaux}, \citenamefont {Lekeu},\ and\ \citenamefont
  {Leonard}}]{Henneaux:2019zod}%
  \BibitemOpen
  \bibfield  {author} {\bibinfo {author} {\bibfnamefont {M.}~\bibnamefont
  {Henneaux}}, \bibinfo {author} {\bibfnamefont {V.}~\bibnamefont {Lekeu}}, \
  and\ \bibinfo {author} {\bibfnamefont {A.}~\bibnamefont {Leonard}},\ }\href
  {\doibase 10.1088/1751-8121/ab56ed} {\bibfield  {journal} {\bibinfo
  {journal} {J. Phys. A}\ }\textbf {\bibinfo {volume} {53}},\ \bibinfo {pages}
  {014002} (\bibinfo {year} {2020})},\ \Eprint
  {http://arxiv.org/abs/1909.12706} {arXiv:1909.12706 [hep-th]} \BibitemShut
  {NoStop}%
\bibitem [{Note1()}]{Note1}%
  \BibitemOpen
  \bibinfo {note} {Although charged under spacetime symmetries, these labels do
  not transform as Lorentz tensors \cite {Benedetti:2022zbb}.}\BibitemShut
  {Stop}%
\bibitem [{\citenamefont {{Arnowitt}}\ \emph {et~al.}(1959)\citenamefont
  {{Arnowitt}}, \citenamefont {{Deser}},\ and\ \citenamefont
  {{Misner}}}]{1959PhRv..116.1322A}%
  \BibitemOpen
  \bibfield  {author} {\bibinfo {author} {\bibfnamefont {R.}~\bibnamefont
  {{Arnowitt}}}, \bibinfo {author} {\bibfnamefont {S.}~\bibnamefont {{Deser}}},
  \ and\ \bibinfo {author} {\bibfnamefont {C.~W.}\ \bibnamefont {{Misner}}},\
  }\href {\doibase 10.1103/PhysRev.116.1322} {\bibfield  {journal} {\bibinfo
  {journal} {Physical Review}\ }\textbf {\bibinfo {volume} {116}},\ \bibinfo
  {pages} {1322} (\bibinfo {year} {1959})}\BibitemShut {NoStop}%
\bibitem [{\citenamefont {Bueno}\ \emph {et~al.}(2017)\citenamefont {Bueno},
  \citenamefont {Cano}, \citenamefont {Min},\ and\ \citenamefont
  {Visser}}]{Bueno:2016ypa}%
  \BibitemOpen
  \bibfield  {author} {\bibinfo {author} {\bibfnamefont {P.}~\bibnamefont
  {Bueno}}, \bibinfo {author} {\bibfnamefont {P.~A.}\ \bibnamefont {Cano}},
  \bibinfo {author} {\bibfnamefont {V.~S.}\ \bibnamefont {Min}}, \ and\
  \bibinfo {author} {\bibfnamefont {M.~R.}\ \bibnamefont {Visser}},\ }\href
  {\doibase 10.1103/PhysRevD.95.044010} {\bibfield  {journal} {\bibinfo
  {journal} {Phys. Rev. D}\ }\textbf {\bibinfo {volume} {95}},\ \bibinfo
  {pages} {044010} (\bibinfo {year} {2017})},\ \Eprint
  {http://arxiv.org/abs/1610.08519} {arXiv:1610.08519 [hep-th]} \BibitemShut
  {NoStop}%
\bibitem [{\citenamefont {Alvarez-Gaume}\ \emph {et~al.}(2016)\citenamefont
  {Alvarez-Gaume}, \citenamefont {Kehagias}, \citenamefont {Kounnas},
  \citenamefont {L\"ust},\ and\ \citenamefont
  {Riotto}}]{Alvarez-Gaume:2015rwa}%
  \BibitemOpen
  \bibfield  {author} {\bibinfo {author} {\bibfnamefont {L.}~\bibnamefont
  {Alvarez-Gaume}}, \bibinfo {author} {\bibfnamefont {A.}~\bibnamefont
  {Kehagias}}, \bibinfo {author} {\bibfnamefont {C.}~\bibnamefont {Kounnas}},
  \bibinfo {author} {\bibfnamefont {D.}~\bibnamefont {L\"ust}}, \ and\ \bibinfo
  {author} {\bibfnamefont {A.}~\bibnamefont {Riotto}},\ }\href {\doibase
  10.1002/prop.201500100} {\bibfield  {journal} {\bibinfo  {journal} {Fortsch.
  Phys.}\ }\textbf {\bibinfo {volume} {64}},\ \bibinfo {pages} {176} (\bibinfo
  {year} {2016})},\ \Eprint {http://arxiv.org/abs/1505.07657} {arXiv:1505.07657
  [hep-th]} \BibitemShut {NoStop}%
\bibitem [{\citenamefont {Stelle}(1977)}]{Stelle:1976gc}%
  \BibitemOpen
  \bibfield  {author} {\bibinfo {author} {\bibfnamefont {K.~S.}\ \bibnamefont
  {Stelle}},\ }\href {\doibase 10.1103/PhysRevD.16.953} {\bibfield  {journal}
  {\bibinfo  {journal} {Phys. Rev. D}\ }\textbf {\bibinfo {volume} {16}},\
  \bibinfo {pages} {953} (\bibinfo {year} {1977})}\BibitemShut {NoStop}%
\bibitem [{\citenamefont {Stelle}(1978)}]{Stelle:1977ry}%
  \BibitemOpen
  \bibfield  {author} {\bibinfo {author} {\bibfnamefont {K.~S.}\ \bibnamefont
  {Stelle}},\ }\href {\doibase 10.1007/BF00760427} {\bibfield  {journal}
  {\bibinfo  {journal} {Gen. Rel. Grav.}\ }\textbf {\bibinfo {volume} {9}},\
  \bibinfo {pages} {353} (\bibinfo {year} {1978})}\BibitemShut {NoStop}%
\bibitem [{\citenamefont {Maldacena}(1998)}]{Maldacena:1997re}%
  \BibitemOpen
  \bibfield  {author} {\bibinfo {author} {\bibfnamefont {J.~M.}\ \bibnamefont
  {Maldacena}},\ }\href {\doibase 10.1023/A:1026654312961} {\bibfield
  {journal} {\bibinfo  {journal} {Adv. Theor. Math. Phys.}\ }\textbf {\bibinfo
  {volume} {2}},\ \bibinfo {pages} {231} (\bibinfo {year} {1998})},\ \Eprint
  {http://arxiv.org/abs/hep-th/9711200} {arXiv:hep-th/9711200} \BibitemShut
  {NoStop}%
\bibitem [{\citenamefont {Aharony}\ \emph {et~al.}(2000)\citenamefont
  {Aharony}, \citenamefont {Gubser}, \citenamefont {Maldacena}, \citenamefont
  {Ooguri},\ and\ \citenamefont {Oz}}]{Aharony_2000}%
  \BibitemOpen
  \bibfield  {author} {\bibinfo {author} {\bibfnamefont {O.}~\bibnamefont
  {Aharony}}, \bibinfo {author} {\bibfnamefont {S.~S.}\ \bibnamefont {Gubser}},
  \bibinfo {author} {\bibfnamefont {J.}~\bibnamefont {Maldacena}}, \bibinfo
  {author} {\bibfnamefont {H.}~\bibnamefont {Ooguri}}, \ and\ \bibinfo {author}
  {\bibfnamefont {Y.}~\bibnamefont {Oz}},\ }\href {\doibase
  10.1016/s0370-1573(99)00083-6} {\bibfield  {journal} {\bibinfo  {journal}
  {Physics Reports}\ }\textbf {\bibinfo {volume} {323}},\ \bibinfo {pages}
  {183–386} (\bibinfo {year} {2000})}\BibitemShut {NoStop}%
\bibitem [{\citenamefont {Weinberg}\ and\ \citenamefont
  {Witten}(1980)}]{WEINBERG198059}%
  \BibitemOpen
  \bibfield  {author} {\bibinfo {author} {\bibfnamefont {S.}~\bibnamefont
  {Weinberg}}\ and\ \bibinfo {author} {\bibfnamefont {E.}~\bibnamefont
  {Witten}},\ }\href {\doibase https://doi.org/10.1016/0370-2693(80)90212-9}
  {\bibfield  {journal} {\bibinfo  {journal} {Physics Letters B}\ }\textbf
  {\bibinfo {volume} {96}},\ \bibinfo {pages} {59} (\bibinfo {year}
  {1980})}\BibitemShut {NoStop}%
\bibitem [{\citenamefont {Pretko}(2017{\natexlab{b}})}]{Pretko:20161}%
  \BibitemOpen
  \bibfield  {author} {\bibinfo {author} {\bibfnamefont {M.}~\bibnamefont
  {Pretko}},\ }\href {\doibase 10.1103/PhysRevB.95.115139} {\bibfield
  {journal} {\bibinfo  {journal} {Phys. Rev. B}\ }\textbf {\bibinfo {volume}
  {95}},\ \bibinfo {pages} {115139} (\bibinfo {year} {2017}{\natexlab{b}})},\
  \Eprint {http://arxiv.org/abs/1604.05329} {arXiv:1604.05329
  [cond-mat.str-el]} \BibitemShut {NoStop}%
\bibitem [{\citenamefont {Pretko}(2017{\natexlab{c}})}]{Pretko:20162}%
  \BibitemOpen
  \bibfield  {author} {\bibinfo {author} {\bibfnamefont {M.}~\bibnamefont
  {Pretko}},\ }\href {\doibase 10.1103/PhysRevB.96.035119} {\bibfield
  {journal} {\bibinfo  {journal} {Phys. Rev. B}\ }\textbf {\bibinfo {volume}
  {96}},\ \bibinfo {pages} {035119} (\bibinfo {year} {2017}{\natexlab{c}})},\
  \Eprint {http://arxiv.org/abs/1606.08857} {arXiv:1606.08857
  [cond-mat.str-el]} \BibitemShut {NoStop}%
\bibitem [{\citenamefont {Blasi}\ and\ \citenamefont
  {Maggiore}(2022)}]{Blasi:2022mbl}%
  \BibitemOpen
  \bibfield  {author} {\bibinfo {author} {\bibfnamefont {A.}~\bibnamefont
  {Blasi}}\ and\ \bibinfo {author} {\bibfnamefont {N.}~\bibnamefont
  {Maggiore}},\ }\href {\doibase 10.1016/j.physletb.2022.137304} {\bibfield
  {journal} {\bibinfo  {journal} {Phys. Lett. B}\ }\textbf {\bibinfo {volume}
  {833}},\ \bibinfo {pages} {137304} (\bibinfo {year} {2022})},\ \Eprint
  {http://arxiv.org/abs/2207.05956} {arXiv:2207.05956 [hep-th]} \BibitemShut
  {NoStop}%
\bibitem [{\citenamefont {Bertolini}\ \emph {et~al.}(2023)\citenamefont
  {Bertolini}, \citenamefont {Blasi}, \citenamefont {Damonte},\ and\
  \citenamefont {Maggiore}}]{Bertolini:2023juh}%
  \BibitemOpen
  \bibfield  {author} {\bibinfo {author} {\bibfnamefont {E.}~\bibnamefont
  {Bertolini}}, \bibinfo {author} {\bibfnamefont {A.}~\bibnamefont {Blasi}},
  \bibinfo {author} {\bibfnamefont {A.}~\bibnamefont {Damonte}}, \ and\
  \bibinfo {author} {\bibfnamefont {N.}~\bibnamefont {Maggiore}},\ }\href
  {\doibase 10.3390/sym15040945} {\bibfield  {journal} {\bibinfo  {journal}
  {Symmetry}\ }\textbf {\bibinfo {volume} {15}},\ \bibinfo {pages} {945}
  (\bibinfo {year} {2023})},\ \Eprint {http://arxiv.org/abs/2304.10789}
  {arXiv:2304.10789 [hep-th]} \BibitemShut {NoStop}%
\bibitem [{\citenamefont {Russo}\ and\ \citenamefont
  {Townsend}(2022)}]{Russo:2022qvz}%
  \BibitemOpen
  \bibfield  {author} {\bibinfo {author} {\bibfnamefont {J.~G.}\ \bibnamefont
  {Russo}}\ and\ \bibinfo {author} {\bibfnamefont {P.~K.}\ \bibnamefont
  {Townsend}},\ }\href@noop {} {\  (\bibinfo {year} {2022})},\ \Eprint
  {http://arxiv.org/abs/2211.10689} {arXiv:2211.10689 [hep-th]} \BibitemShut
  {NoStop}%
\bibitem [{\citenamefont {Guerrieri}\ \emph {et~al.}(2022)\citenamefont
  {Guerrieri}, \citenamefont {Murali}, \citenamefont {Penedones},\ and\
  \citenamefont {Vieira}}]{Guerrieri:2022sod}%
  \BibitemOpen
  \bibfield  {author} {\bibinfo {author} {\bibfnamefont {A.}~\bibnamefont
  {Guerrieri}}, \bibinfo {author} {\bibfnamefont {H.}~\bibnamefont {Murali}},
  \bibinfo {author} {\bibfnamefont {J.}~\bibnamefont {Penedones}}, \ and\
  \bibinfo {author} {\bibfnamefont {P.}~\bibnamefont {Vieira}},\ }\href@noop {}
  {\  (\bibinfo {year} {2022})},\ \Eprint {http://arxiv.org/abs/2212.00151}
  {arXiv:2212.00151 [hep-th]} \BibitemShut {NoStop}%
\bibitem [{\citenamefont {Benedetti}\ \emph
  {et~al.}(2022{\natexlab{c}})\citenamefont {Benedetti}, \citenamefont
  {Casini},\ and\ \citenamefont {Magan}}]{Benedetti:2022ofj}%
  \BibitemOpen
  \bibfield  {author} {\bibinfo {author} {\bibfnamefont {V.}~\bibnamefont
  {Benedetti}}, \bibinfo {author} {\bibfnamefont {H.}~\bibnamefont {Casini}}, \
  and\ \bibinfo {author} {\bibfnamefont {J.~M.}\ \bibnamefont {Magan}},\
  }\href@noop {} {\  (\bibinfo {year} {2022}{\natexlab{c}})},\ \Eprint
  {http://arxiv.org/abs/2212.11291} {arXiv:2212.11291 [hep-th]} \BibitemShut
  {NoStop}%
\bibitem [{\citenamefont {Cano}\ and\ \citenamefont
  {Murcia}(2021)}]{Cano:2021tfs}%
  \BibitemOpen
  \bibfield  {author} {\bibinfo {author} {\bibfnamefont {P.~A.}\ \bibnamefont
  {Cano}}\ and\ \bibinfo {author} {\bibfnamefont {A.}~\bibnamefont {Murcia}},\
  }\href {\doibase 10.1007/JHEP08(2021)042} {\bibfield  {journal} {\bibinfo
  {journal} {JHEP}\ }\textbf {\bibinfo {volume} {08}},\ \bibinfo {pages} {042}
  (\bibinfo {year} {2021})},\ \Eprint {http://arxiv.org/abs/2104.07674}
  {arXiv:2104.07674 [hep-th]} \BibitemShut {NoStop}%
\bibitem [{\citenamefont {Denisov}\ \emph {et~al.}(2017)\citenamefont
  {Denisov}, \citenamefont {Dolgaya}, \citenamefont {Sokolov},\ and\
  \citenamefont {Denisova}}]{Denisov:2017qou}%
  \BibitemOpen
  \bibfield  {author} {\bibinfo {author} {\bibfnamefont {V.~I.}\ \bibnamefont
  {Denisov}}, \bibinfo {author} {\bibfnamefont {E.~E.}\ \bibnamefont
  {Dolgaya}}, \bibinfo {author} {\bibfnamefont {V.~A.}\ \bibnamefont
  {Sokolov}}, \ and\ \bibinfo {author} {\bibfnamefont {I.~P.}\ \bibnamefont
  {Denisova}},\ }\href {\doibase 10.1103/PhysRevD.96.036008} {\bibfield
  {journal} {\bibinfo  {journal} {Phys. Rev. D}\ }\textbf {\bibinfo {volume}
  {96}},\ \bibinfo {pages} {036008} (\bibinfo {year} {2017})}\BibitemShut
  {NoStop}%
\bibitem [{\citenamefont {Sorokin}(2022)}]{Sorokin:2021tge}%
  \BibitemOpen
  \bibfield  {author} {\bibinfo {author} {\bibfnamefont {D.~P.}\ \bibnamefont
  {Sorokin}},\ }\href {\doibase 10.1002/prop.202200092} {\bibfield  {journal}
  {\bibinfo  {journal} {Fortsch. Phys.}\ }\textbf {\bibinfo {volume} {70}},\
  \bibinfo {pages} {2200092} (\bibinfo {year} {2022})},\ \Eprint
  {http://arxiv.org/abs/2112.12118} {arXiv:2112.12118 [hep-th]} \BibitemShut
  {NoStop}%
\bibitem [{\citenamefont {Born}\ and\ \citenamefont
  {Infeld}(1934)}]{Born:1934gh}%
  \BibitemOpen
  \bibfield  {author} {\bibinfo {author} {\bibfnamefont {M.}~\bibnamefont
  {Born}}\ and\ \bibinfo {author} {\bibfnamefont {L.}~\bibnamefont {Infeld}},\
  }\href {\doibase 10.1098/rspa.1934.0059} {\bibfield  {journal} {\bibinfo
  {journal} {Proc. Roy. Soc. Lond. A}\ }\textbf {\bibinfo {volume} {144}},\
  \bibinfo {pages} {425} (\bibinfo {year} {1934})}\BibitemShut {NoStop}%
\bibitem [{\citenamefont {Bandos}\ \emph {et~al.}(2020)\citenamefont {Bandos},
  \citenamefont {Lechner}, \citenamefont {Sorokin},\ and\ \citenamefont
  {Townsend}}]{Bandos:2020jsw}%
  \BibitemOpen
  \bibfield  {author} {\bibinfo {author} {\bibfnamefont {I.}~\bibnamefont
  {Bandos}}, \bibinfo {author} {\bibfnamefont {K.}~\bibnamefont {Lechner}},
  \bibinfo {author} {\bibfnamefont {D.}~\bibnamefont {Sorokin}}, \ and\
  \bibinfo {author} {\bibfnamefont {P.~K.}\ \bibnamefont {Townsend}},\ }\href
  {\doibase 10.1103/PhysRevD.102.121703} {\bibfield  {journal} {\bibinfo
  {journal} {Phys. Rev. D}\ }\textbf {\bibinfo {volume} {102}},\ \bibinfo
  {pages} {121703} (\bibinfo {year} {2020})},\ \Eprint
  {http://arxiv.org/abs/2007.09092} {arXiv:2007.09092 [hep-th]} \BibitemShut
  {NoStop}%
\bibitem [{\citenamefont {Bandos}\ \emph {et~al.}(2021)\citenamefont {Bandos},
  \citenamefont {Lechner}, \citenamefont {Sorokin},\ and\ \citenamefont
  {Townsend}}]{Bandos:2021rqy}%
  \BibitemOpen
  \bibfield  {author} {\bibinfo {author} {\bibfnamefont {I.}~\bibnamefont
  {Bandos}}, \bibinfo {author} {\bibfnamefont {K.}~\bibnamefont {Lechner}},
  \bibinfo {author} {\bibfnamefont {D.}~\bibnamefont {Sorokin}}, \ and\
  \bibinfo {author} {\bibfnamefont {P.~K.}\ \bibnamefont {Townsend}},\ }\href
  {\doibase 10.1007/JHEP10(2021)031} {\bibfield  {journal} {\bibinfo  {journal}
  {JHEP}\ }\textbf {\bibinfo {volume} {10}},\ \bibinfo {pages} {031} (\bibinfo
  {year} {2021})},\ \Eprint {http://arxiv.org/abs/2106.07547} {arXiv:2106.07547
  [hep-th]} \BibitemShut {NoStop}%
\bibitem [{\citenamefont {Kosyakov}(2020)}]{Kosyakov:2020wxv}%
  \BibitemOpen
  \bibfield  {author} {\bibinfo {author} {\bibfnamefont {B.~P.}\ \bibnamefont
  {Kosyakov}},\ }\href {\doibase 10.1016/j.physletb.2020.135840} {\bibfield
  {journal} {\bibinfo  {journal} {Phys. Lett. B}\ }\textbf {\bibinfo {volume}
  {810}},\ \bibinfo {pages} {135840} (\bibinfo {year} {2020})},\ \Eprint
  {http://arxiv.org/abs/2007.13878} {arXiv:2007.13878 [hep-th]} \BibitemShut
  {NoStop}%
\bibitem [{\citenamefont {Liu}\ \emph {et~al.}(2020)\citenamefont {Liu},
  \citenamefont {Mai}, \citenamefont {Li},\ and\ \citenamefont
  {L\"u}}]{Liu:2019rib}%
  \BibitemOpen
  \bibfield  {author} {\bibinfo {author} {\bibfnamefont {H.-S.}\ \bibnamefont
  {Liu}}, \bibinfo {author} {\bibfnamefont {Z.-F.}\ \bibnamefont {Mai}},
  \bibinfo {author} {\bibfnamefont {Y.-Z.}\ \bibnamefont {Li}}, \ and\ \bibinfo
  {author} {\bibfnamefont {H.}~\bibnamefont {L\"u}},\ }\href {\doibase
  10.1007/s11433-019-1446-1} {\bibfield  {journal} {\bibinfo  {journal} {Sci.
  China Phys. Mech. Astron.}\ }\textbf {\bibinfo {volume} {63}},\ \bibinfo
  {pages} {240411} (\bibinfo {year} {2020})},\ \Eprint
  {http://arxiv.org/abs/1907.10876} {arXiv:1907.10876 [hep-th]} \BibitemShut
  {NoStop}%
\bibitem [{\citenamefont {Cisterna}\ \emph {et~al.}(2020)\citenamefont
  {Cisterna}, \citenamefont {Giribet}, \citenamefont {Oliva},\ and\
  \citenamefont {Pallikaris}}]{Cisterna:2020rkc}%
  \BibitemOpen
  \bibfield  {author} {\bibinfo {author} {\bibfnamefont {A.}~\bibnamefont
  {Cisterna}}, \bibinfo {author} {\bibfnamefont {G.}~\bibnamefont {Giribet}},
  \bibinfo {author} {\bibfnamefont {J.}~\bibnamefont {Oliva}}, \ and\ \bibinfo
  {author} {\bibfnamefont {K.}~\bibnamefont {Pallikaris}},\ }\href {\doibase
  10.1103/PhysRevD.101.124041} {\bibfield  {journal} {\bibinfo  {journal}
  {Phys. Rev. D}\ }\textbf {\bibinfo {volume} {101}},\ \bibinfo {pages}
  {124041} (\bibinfo {year} {2020})},\ \Eprint
  {http://arxiv.org/abs/2004.05474} {arXiv:2004.05474 [hep-th]} \BibitemShut
  {NoStop}%
\bibitem [{\citenamefont {Casini}\ \emph {et~al.}(2022)\citenamefont {Casini},
  \citenamefont {Magan},\ and\ \citenamefont {Martinez}}]{Pedro}%
  \BibitemOpen
  \bibfield  {author} {\bibinfo {author} {\bibfnamefont {H.}~\bibnamefont
  {Casini}}, \bibinfo {author} {\bibfnamefont {J.~M.}\ \bibnamefont {Magan}}, \
  and\ \bibinfo {author} {\bibfnamefont {P.~J.}\ \bibnamefont {Martinez}},\
  }\href {\doibase 10.1007/JHEP01(2022)079} {\bibfield  {journal} {\bibinfo
  {journal} {JHEP}\ }\textbf {\bibinfo {volume} {01}},\ \bibinfo {pages} {079}
  (\bibinfo {year} {2022})},\ \Eprint {http://arxiv.org/abs/2110.02980}
  {arXiv:2110.02980 [hep-th]} \BibitemShut {NoStop}%
\bibitem [{\citenamefont {Sisman}\ \emph {et~al.}(2011)\citenamefont {Sisman},
  \citenamefont {Gullu},\ and\ \citenamefont {Tekin}}]{Sisman:2011gz}%
  \BibitemOpen
  \bibfield  {author} {\bibinfo {author} {\bibfnamefont {T.~C.}\ \bibnamefont
  {Sisman}}, \bibinfo {author} {\bibfnamefont {I.}~\bibnamefont {Gullu}}, \
  and\ \bibinfo {author} {\bibfnamefont {B.}~\bibnamefont {Tekin}},\ }\href
  {\doibase 10.1088/0264-9381/28/19/195004} {\bibfield  {journal} {\bibinfo
  {journal} {Class. Quant. Grav.}\ }\textbf {\bibinfo {volume} {28}},\ \bibinfo
  {pages} {195004} (\bibinfo {year} {2011})},\ \Eprint
  {http://arxiv.org/abs/1103.2307} {arXiv:1103.2307 [hep-th]} \BibitemShut
  {NoStop}%
\bibitem [{\citenamefont {Lu}\ and\ \citenamefont {Pope}(2011)}]{Lu:2011zk}%
  \BibitemOpen
  \bibfield  {author} {\bibinfo {author} {\bibfnamefont {H.}~\bibnamefont
  {Lu}}\ and\ \bibinfo {author} {\bibfnamefont {C.~N.}\ \bibnamefont {Pope}},\
  }\href {\doibase 10.1103/PhysRevLett.106.181302} {\bibfield  {journal}
  {\bibinfo  {journal} {Phys. Rev. Lett.}\ }\textbf {\bibinfo {volume} {106}},\
  \bibinfo {pages} {181302} (\bibinfo {year} {2011})},\ \Eprint
  {http://arxiv.org/abs/1101.1971} {arXiv:1101.1971 [hep-th]} \BibitemShut
  {NoStop}%
\bibitem [{\citenamefont {Bueno}\ and\ \citenamefont
  {Cano}(2016)}]{Bueno:2016xff}%
  \BibitemOpen
  \bibfield  {author} {\bibinfo {author} {\bibfnamefont {P.}~\bibnamefont
  {Bueno}}\ and\ \bibinfo {author} {\bibfnamefont {P.~A.}\ \bibnamefont
  {Cano}},\ }\href {\doibase 10.1103/PhysRevD.94.104005} {\bibfield  {journal}
  {\bibinfo  {journal} {Phys. Rev. D}\ }\textbf {\bibinfo {volume} {94}},\
  \bibinfo {pages} {104005} (\bibinfo {year} {2016})},\ \Eprint
  {http://arxiv.org/abs/1607.06463} {arXiv:1607.06463 [hep-th]} \BibitemShut
  {NoStop}%
\bibitem [{\citenamefont {Salvio}(2018)}]{Salvio:2018crh}%
  \BibitemOpen
  \bibfield  {author} {\bibinfo {author} {\bibfnamefont {A.}~\bibnamefont
  {Salvio}},\ }\href {\doibase 10.3389/fphy.2018.00077} {\bibfield  {journal}
  {\bibinfo  {journal} {Front. in Phys.}\ }\textbf {\bibinfo {volume} {6}},\
  \bibinfo {pages} {77} (\bibinfo {year} {2018})},\ \Eprint
  {http://arxiv.org/abs/1804.09944} {arXiv:1804.09944 [hep-th]} \BibitemShut
  {NoStop}%
\bibitem [{\citenamefont {Lovelock}(1971)}]{Lovelock:1971yv}%
  \BibitemOpen
  \bibfield  {author} {\bibinfo {author} {\bibfnamefont {D.}~\bibnamefont
  {Lovelock}},\ }\href {\doibase 10.1063/1.1665613} {\bibfield  {journal}
  {\bibinfo  {journal} {J. Math. Phys.}\ }\textbf {\bibinfo {volume} {12}},\
  \bibinfo {pages} {498} (\bibinfo {year} {1971})}\BibitemShut {NoStop}%
\bibitem [{\citenamefont {Padmanabhan}\ and\ \citenamefont
  {Kothawala}(2013)}]{Padmanabhan:2013xyr}%
  \BibitemOpen
  \bibfield  {author} {\bibinfo {author} {\bibfnamefont {T.}~\bibnamefont
  {Padmanabhan}}\ and\ \bibinfo {author} {\bibfnamefont {D.}~\bibnamefont
  {Kothawala}},\ }\href {\doibase 10.1016/j.physrep.2013.05.007} {\bibfield
  {journal} {\bibinfo  {journal} {Phys. Rept.}\ }\textbf {\bibinfo {volume}
  {531}},\ \bibinfo {pages} {115} (\bibinfo {year} {2013})},\ \Eprint
  {http://arxiv.org/abs/1302.2151} {arXiv:1302.2151 [gr-qc]} \BibitemShut
  {NoStop}%
\bibitem [{\citenamefont {Landau}(1980)}]{alma991009084709703276}%
  \BibitemOpen
  \bibfield  {author} {\bibinfo {author} {\bibfnamefont {L.~D.}\ \bibnamefont
  {Landau}},\ }\href@noop {} {\emph {\bibinfo {title} {Statistical physics by
  L.D. Landau and E.M. Lifshitz ; translated from the Russian by J.R. Sykes and
  M.J. Kearsley}}},\ \bibinfo {edition} {3rd}\ ed.,\ Their Course of
  theoretical physics , v. 9, pt. 2\ (\bibinfo  {publisher} {Pergamon Press},\
  \bibinfo {address} {Oxford},\ \bibinfo {year} {1980})\BibitemShut {NoStop}%
\end{thebibliography}%
\bibliographystyle{apsrev4-1}
\end{document}